\documentclass[%
 reprint,
superscriptaddress,
 amsmath,amssymb,
 aps,
 prl,
]{revtex4-2}

\usepackage{graphicx}
\usepackage{dcolumn}
\usepackage{bm}
\PassOptionsToPackage{hyphens}{url}
\usepackage[hidelinks]{hyperref}
\hypersetup{breaklinks=true}

\usepackage{enumitem}
\usepackage{xcolor}
\usepackage{caption}
\usepackage{subcaption}
\usepackage{lipsum}
\usepackage{pgfplots}

\definecolor{tissueG}{HTML}{B3FFB3}
\definecolor{nuclei}{HTML}{910f00}
\usetikzlibrary{backgrounds}
\pgfplotsset{compat=1.18}
\usetikzlibrary {arrows.meta}

\newcommand{\bv}[1]{\mathbf{{#1}}}

\newcommand{\uVec}{\hat{\mathrm{e}}}

\newcommand{\curl}{\nabla \times}

\newcommand{\funcdiff}[2]{\frac{\delta #1}{\delta #2}}
\newcommand{\dd}{\mathrm{d}}

\newcommand{\order}{\mathcal{O}}

\newcommand{\trace}{\mathrm{Tr}}

\newcommand{\pert}{\epsilon_\alpha}

\begin{document}


\title{Dynamics of Wound Closure in Living Nematic Epithelia}

\author{Henry Andralojc}
\email{h.andralojc@bristol.ac.uk}
\affiliation{School of Mathematics, University of Bristol, Fry Building, Woodland Road, Bristol BS8 1UG, UK}
\author{Jake Turley}
\affiliation{School of Mathematics, University of Bristol, Fry Building, Woodland Road, Bristol BS8 1UG, UK}
\affiliation{Mechanobiology Institute, National University of Singapore, Singapore, Singapore}
\affiliation{School of Biochemistry, University of Bristol, University Walk, Bristol, BS8 1TD, UK}
\author{Helen Weavers}
\affiliation{School of Biochemistry, University of Bristol, University Walk, Bristol, BS8 1TD, UK}
\author{Paul Martin}
\affiliation{School of Biochemistry, University of Bristol, University Walk, Bristol, BS8 1TD, UK}
\author{Isaac V. Chenchiah}
\affiliation{School of Mathematics, University of Bristol, Fry Building, Woodland Road, Bristol BS8 1UG, UK}
\author{Rachel R. Bennett}
\email{rachel.bennett@bristol.ac.uk}
\affiliation{School of Mathematics, University of Bristol, Fry Building, Woodland Road, Bristol BS8 1UG, UK}
\affiliation{Isaac Newton Institute for
Mathematical Sciences, 20 Clarkson Rd, Cambridge CB3 0EH, UK}
\author{Tanniemola B. Liverpool}
\email{t.liverpool@bristol.ac.uk}
\affiliation{School of Mathematics, University of Bristol, Fry Building, Woodland Road, Bristol BS8 1UG, UK}
\affiliation{Isaac Newton Institute for
Mathematical Sciences, 20 Clarkson Rd, Cambridge CB3 0EH, UK}

\date{\today}

\begin{abstract}
We study theoretically the closure of a wound in a layer of epithelial cells in a living tissue after damage.  Our analysis is informed by our recent experiments observing  re-epithelialisation \textit{in vivo} of \textit{Drosophila} pupae. On time and length-scales such that the evolution of the epithelial tissue near the wound is well captured by that of a 2D active fluid with local nematic order, we consider the free-surface problem of a hole in a bounded region of tissue, and study the role that active stresses far from the hole play in the closure of the hole.  For parallel anchored nematic order  at the wound boundary (as we observe in our experiments), we find that closure is accelerated when the active stresses are contractile and slowed down when the stresses are extensile. Parallel anchoring also leads to the appearance of topological defects which annihilate upon wound closure.
\end{abstract}

\maketitle


Tissue damage triggers a complex series of overlapping cell and tissue movements that are reminiscent of many of the processes found in embryonic morphogenesis \cite{collinet_programmed_2021}. These processes together stave off infection and eventually repair the wound, bringing the tissue back to something approaching its pre-wounded state \cite{gurtner_wound_2008, eming_wound_2014, pena_cellular_2024, degen_embryonic_2012, martin_cellular_2015}. A key stage of wound healing is  re-epithelialisation, in which the epidermal cells at the cut wound edge and the sheet of epithelium behind them advance to seal the wound gap. Individual and concerted cell contractions and {\em shape changes}, as well as cell {\em movements} and/or migration and cell {\em divisions}, all contribute to the closure of the epithelial gap \cite{park_tissue-scale_2017, aragona_defining_2017, razzell_recapitulation_2014, tetley_tissue_2019, turley_deep_2024-1, garcia-fernandez_epithelial_2009}. However precisely how much each of these different cell behaviours, and the forces associated with them, contributes to the ultimate wound healing goal is not known. This is because tissue in many organisms are opaque, and hence it is difficult to quantify the contribution of these processes. Therefore the study of translucent tissue from genetically tractable organisms like {\em Drosophila melanogaster}~\cite{wood_wound_2002,galko_cellular_2004,razzell_recapitulation_2014,weavers_systems_2016, ramet_jnk_2002, franz_fat_2018, mccluskey_study_1993}, can play a valuable role in leading us towards the required  mechanistic understanding of the process \cite{razzell_swatting_2011,munoz-soriano_why_2014, turley_what_2022}.  The hope is that studying the cellular responses to wounding in such model systems can give insight into mammalian healing, ultimately leading to the development of practices that can aid clinicians and patients \cite{sonnemann_wound_2011,degen_embryonic_2012}.  Such studies can also inspire design principles for developing self-healing artificial systems.

\begin{figure*}
\input{./figures/fig1new.tex}
    \caption{\raggedright (a): Time-lapse movies of the developing \textit{Drosophila} pupal epithelium (18 hrs APF) indicate that unwounded tissue has nematic order. Elongation and alignment of each epithelial cell is described by a traceless shape tensor, $q$, the magnitude of which $||q|| = \frac{1}{2} \trace (q^2)$ describes anisotropy in shape and $\phi \in [-\pi/2, \pi/2)$ describes the direction of nematic alignment relative to the $x$-axis. (b) Epithelial cells align, on average, along the PD axis of the wing (roughly aligning with the $x$-axis of our imaging setup), the degree of elongation increasing as the tissue develops. (c): Snapshot of tissue $\sim 8$ minutes post wounding. Here, the cellular alignment, $\phi^\prime$ is measured relative to the radial direction in a polar basis centred on the wound as illustrated in the inset: red (blue) cells indicate alignment along $\pm \uVec_\theta$ ($\pm \uVec_r$). (d): Distribution of $|\phi^\prime|$'s for cells inside annulus illustrated in panel (c). We find that cells close to the wound initially tend to be aligned along the $\pm \uVec_\theta$ direction. (e): Illustration of the model setup and (f): the active contractile and extensile force dipoles included in the model that coarse-grain to generate a stress $\sim \alpha \bv{Q}$.}
\label{fig:modelSetup}
\end{figure*}

Continuum models for re-epithelialisation typically consider the surrounding tissue to be an isotropic medium where closure is driven by proliferation/growth processes \cite{arciero_continuum_2011, bowden_morphoelastic_2016} and/or actin-based machineries localised at the wound edge \cite{poujade_collective_2007, cochet-escartin_border_2014, almeida_modeling_2012, ravasio_gap_2015, arciero_continuum_2011, chen_free_2000, xue_mathematical_2009, mark_physical_2010}, and with the possibility of coupling between mechanical stresses and chemical signals \cite{roldan_computational_2019} in the tissue. The effects of the actin `purse-string' \cite{martin_actin_1992, wood_wound_2002, kamran_vivo_2017, tamada_two_2007, martin_repair_1994} (a bundle of filaments at the wound edge that constricts, dragging the surrounding tissue radially inwards) and filopodia/lamellipodia (motile extrusions from the leading edge cells that extend into the wound space and drag the cells behind \cite{brugues_forces_2014}) can both be modelled by including additional (active) forces in the boundary conditions at the wound surface. It is believed, however, that mechanical forces in cells far from the edge also have a role to play in efficient epithelial gap/wound closure \cite{razzell_recapitulation_2014, matsubayashi_white_2011, antunes_coordinated_2013, tetley_tissue_2019, park_tissue-scale_2017, lee_crawling_2011}. Therefore, in this work, we present a general theoretical continuum framework for re-epithelialisation that includes active forces generated in the surrounding bulk tissue \cite{villeneuve_measuring_2025, ladoux_mechanobiology_2017, brugues_forces_2014, tetley_tissue_2019, etournay_interplay_2015, comelles_epithelial_2021,
schotz_glassy_2013}.

Inspired by our recent experiments observing wound healing in the developing \textit{Drosophila} pupal wing epithelium~\cite{olenik_fluctuations_2023, turley_deep_2024,turley_deep_2024-1, turley_deep_2023}, we model the epithelium as an incompressible, two-dimensional active nematic and study the closure of a hole (model wound) in this continuum fluid. Focusing on the effect of mechanical activity in the bulk tissue on wound closure, we find that for parallel anchoring of the nematic order at the wound boundary and $x$-axis alignment in the far-field -- consistent with cellular alignments observed in our experiments -- contractile active stresses accelerate wound closure, whilst extensile ones delay it. Our model also reproduces the anisotropy in wound shapes observed in experiment, highlighting the importance of bulk active forces in the modelling of epithelial closure events. 

This paper is outlined as follows. First, we present some experimental analysis of the cell shapes and alignments in unwounded and wounded tissue which justify the nematic description and the choices of boundary conditions in our model. Next we present our continuum model and outline some key results (full calculational details for which can be found in the SI). Finally we make connection again with our experiments by presenting results of analysis of the anisotropy in observed wound shapes and compare with the predictions of our model.

\paragraph*{\label{sec:experiment}Experimental Analysis of Cell Shapes.} Our wounding experiments used laser ablation to create an initially approximately circular wound in the wing epithelium of developing \textit{Drosophila} pupae which was tracked using time-lapse confocal microscopy \cite{weavers_long-term_2018}. Cell shapes, motion and divisions were then tracked using flourescent tagging and machine learning~\cite{olenik_fluctuations_2023,turley_deep_2023}. Following segmentation, the cell and wound boundaries were approximated by polygons and moments over each polygons shape were calculated to quantify locations, areas and shapes of the features in each time-lapse. Of particular interest to us here is the traceless shape tensor, $q$, which quantifies the degree of elongation and alignment of a polygon, defined as:
\begin{equation}
q = \frac{1}{A^2} \int_\mathrm{poly} \dd x \dd y \begin{pmatrix}
    x^2-y^2 & xy \\
    xy & y^2 - x^2
\end{pmatrix}\, ,
\end{equation}

where the integral is taken over the polygon and $A$ is the area of the polygon in question. Analogous to the rank-2 nematic order parameter \cite{de_gennes_short_1971}, $q$ may be written in terms of alignments along a basis set, e.g. $q = \big(\begin{smallmatrix}
  q_{xx} & q_{xy} \\
  q_{xy} & - q_{xx}
\end{smallmatrix}\big)$, or in terms of a magnitude and angle $q = ||q|| \big(\begin{smallmatrix}
  \cos 2\phi & \sin 2\phi \\
  \sin 2\phi & -\cos 2\phi
\end{smallmatrix}\big)$, where the magnitude $||q|| = \frac{1}{2} \trace (q^2) = q_{xx}^2 + q_{xy}^2$ gives a measure for the degree of elongation of each polygon ($||q|| = 0$ for circular polygons, $||q||\neq 0$ for anisotropic polygons), and the angle, defined as $\tan 2 \phi = \frac{q_{xy}}{q_{xx}}$ defines the axis of alignment for the polygon relative to the $x$-axis. We will use $||q||$ as a measure for anisotropy in both cell and wound shapes.

We make a two key observations about cellular shapes. First, in healthy tissue, the typical cell polygon was elongated, head-tail symmetric \cite{turley_deep_2023} and aligned along the long axis of the wing -- that is, the tissue has nematic symmetry. Fig.~\ref{fig:modelSetup}a illustrates a snapshot of the segmented cells in an unwounded sample, color coded by magnitude of $||q||$ and angle relative to the $x$-axis in each half. Averaging over all cells in each frame (Fig.~\ref{fig:modelSetup}b), we find that the typical cell elongates along the proximal-distal axis of the developing wing (roughly aligning with the $x$-axis of our imaging setup). Second, by rotating each cell's $q$ into a polar basis centred on the wound (see SI), we find that the cells close to a wound tend to align not along the bulk tissue axis but parallel to the wound edge (Fig.~\ref{fig:modelSetup}c-d). Finally, we also observed a dramatic decrease in cellular division near wounds following wounding~\cite{turley_deep_2024, turley_deep_2024-1} -- i.e. the tissue is initially approximately incompressible~\cite{cochet-escartin_border_2014}. 

These three observations combined suggest that, in the continuum limit, the wounded epithelium can be modelled as an incompressible, two-dimensional active nematic, \cite{saw_biological_2018, doostmohammadi_active_2018, ramaswamy_mechanics_2010, marchetti_hydrodynamics_2013} with a hole (wound) at its centre with parallel anchoring at the wound edge and $x$-axis alignment in the far-field.

\paragraph*{\label{sec:model}Model.}
Local cell shape anisotropy in our model is described the traceless symmetric $2 \times 2$  tensor field, $\bv{Q} (\bv{r},t)$ \cite{de_gennes_short_1971,chaikin_principles_1995} and motion by an incompressible velocity field, $\bv{v}(\bv{r},t)$ and an associated pressure $p(\bv{r},t)$. We study the time-evolution of these fields in a bounded domain following the removal of a circular disc from its center at $t=0$. On the time-scale of healing in this system (up to $\sim$ 4 hours \cite{turley_deep_2024-1}), we assume the tissue flows like a viscous fluid, but with the possibility of additional active stresses. Therefore we use the equations of active nematohydrodynamics at vanishing Reynolds number in contact with a frictional substrate~\cite{marchetti_hydrodynamics_2013,doostmohammadi_active_2018, beris_thermodynamics_1994, cates_theories_2018}:

\begin{subequations}
\begin{gather}
0 = \nabla \cdot \bv{v} \;, \\
0 = -\nabla p + \eta \nabla^2 \bv{v} - \Gamma \bv{v} + \alpha \nabla \cdot \bv{Q} \;, \\
0 = \funcdiff{\mathcal{F}_\mathrm{LdG}}{\bv{Q}} \;.
\end{gather}
\label{eqn:equationsOfMotion}
\end{subequations}
where  $\bv{\mathcal{F}}_\mathrm{LdG}$ is a Landau-de Gennes free energy \cite{de_gennes_short_1971,marchetti_hydrodynamics_2013}:%
\begin{equation}\label{eqn:LdG}
    \mathcal{F}_{\mathrm{LdG}} = \int \dd x \left[-\frac{A}{2} ||\bv{Q}||^2 + \frac{B}{4} ||\bv{Q}||^4 + \frac{K}{2} ||\nabla \bv{Q}||^2 \right]\;,
\end{equation}
with $||\bv{Q}||^2=Q_{ij}Q_{ij}$, $||\nabla \bv{Q}||^2 = \partial_i Q_{jk} \partial_i Q_{jk}$. 
$\eta$ is the bulk viscosity and $\Gamma$ is the friction coefficient, which we include to model the effect of a basal membrane resisting the flow of the tissue above it. The active stress $\sigma^a_{ij}= \alpha Q_{ij}$ is present when  $\alpha \ne 0$~\cite{simha_hydrodynamic_2002, marchetti_hydrodynamics_2013}. We have assumed that  $\bv{Q}$ relaxes much faster than $\bv{v}$, so the nematic texture is always in a local equilibrium. 
We consider the wound sufficiently far from the outer wing boundary that it does not affect the wound and choose, for simplicity, an initially circular outer boundary (see Fig. \ref{fig:modelSetup}e). We also ignore elastic contributions to the fluid stress arising due to variations in $\bv{Q}$ as they are higher order in gradient and subdominant at long lengthscales~\cite{marchetti_hydrodynamics_2013}. 

We solve these subject to standard kinematic and dynamic boundary conditions on each free surface, labelled by $r=R_i(\theta,t)$ for $i=1,2$ for the inner and outer free surfaces respectively, in addition to conditions on the components of $\bv{Q}$ (see eqn.~SX). The dynamic boundary condition (DBC) imposes force balance on every point on the free boundary:
\begin{equation}\label{eqn:DBCs}
\left[\left(p^\mathrm{ext}_i - p\right) \bv{n} + (2\eta \bv{D} + \alpha \bv{Q}) \cdot \bv{n} \right] \big |_{R_i} = \gamma \kappa_i \bv{n} \;,
\end{equation}
where $p^\mathrm{ext}_i$ is the pressure of the exterior fluid in contact with the $i^\text{th}$ free surface, $D_{jk}=\frac{1}{2}(\partial_j v_k + \partial_k v_j)$ is the rate of strain tensor \cite{batchelor_introduction_2000}, $\bv{n}$ is the unit normal to the boundary and $\kappa_i$ is the local curvature of the $i^\text{th}$ surface. 
The effective surface tension $\gamma = \tilde{\gamma} + \tau > 0 $ contains both a constant passive surface tension $\tilde{\gamma}$ and an active term $\tau$, which we include to model the possible presence of a constricting actin purse string, see SI for justification. 
The kinematic boundary condition (KBC) then governs the time evolution of the free surfaces, requiring that the flow convects the boundary:
\begin{equation}
	D_t \left[r-R_i(\theta,t)\right] = 0  \quad \Rightarrow \quad v_r |_{R_i} = \dot{R}_i + \frac{v_\theta |_{R_i}}{R_i} R^\prime_i \;,
\end{equation}
where $D_t = \partial_t + \bv{v} \cdot \nabla$ is the Lagrangian material derivative. 

We non-dimensionalise our equations: $\bv{r} \to L \bv{r},\, t \to T t,\, p \to \Pi p, \, \bv{Q} \to Q_0 \bv{Q}$ choosing $L =  \sqrt{\eta/\Gamma}, \, T = \eta/\Pi, \, \Pi = \gamma/L, \, Q_0 = \sqrt{A/2B} \, , $ and define non-dimensional parameters $\pert = \alpha Q_0/ \Pi$,  (relative importance of active to other stresses) and $\Lambda = L/\ell_Q$, (relative sizes of the flow and nematic relaxation length, $\ell_Q^2 = K/2A$, scales). $1/\Lambda$ controls the nematic length scale, and thus the persistence of active forces into the bulk of the fluid.

The wounds in our experiments are initially approximately circular \cite{turley_deep_2023}. Therefore, we solve equations~\eqref{eqn:equationsOfMotion} by expanding the dynamical fields in the activity, $\pert$:
\begin{subequations}
\begin{gather}
R_i(\theta,t) = R^0_i(t) + \pert R^1_i(\theta,t) \\
\bv{v} (r,\theta,t) = \bv{v}^0(r,t) + \pert \bv{v}^1(r,\theta,t) + ...\\
p = p^0 + \pert p^1 + ... \\
\bv{Q} = \bv{Q}^0 + \pert \bv{Q}^1 + ... \; , 
\end{gather}
\end{subequations}
and decomposing into Fourier modes~\cite{alert_fingering_2022}:
\begin{equation}
R_i^1(\theta,t) = \sum_{n=0}^\infty \xi_i^n(t) \cos n\theta + \eta_i^n(t) \sin n\theta \;, 
\label{eq:Fourier-mode}\end{equation}
so the leading-order problem is a droplet of passive viscous fluid with a circular hole at the centre, whose closure is driven by an (active) surface tension and differences in external pressures. The flow at first order in activity is additionally driven by leading-order gradients in $\bv{Q}$ -- i.e. the nematic response to a circular boundary. The magnitude of the active stress in tissue typically cannot be measured directly, and so we cannot confidently claim that $\epsilon_\alpha$ is a small parameter, however the fact that the wounds inflicted are initially of circular cross-section suggest it is {\em a posteriori}, $\epsilon_\alpha R_i^1/R_i^0 \ll 1$ for short times. The continuum approach will also of course break down as the wound size approaches the typical cell size and the division rate returns to levels observed in healthy tissue. Therefore, although we solve the equations of motion up to closure, we shall mainly focus on short times following wounding.
\begin{figure*}[t]
\centering
\input{./figures/fig23Combinednew.tex}
\caption{\raggedright (a): Nematic texture (left) and active force (right) that appears in the Stokes equation and drives the flow. The blue squares mark positions of $-\frac{1}{2}$ topological defects. (b): Snapshots of nematic texture throughout closure show the motion of defects towards the origin, resulting in a topologically `healed' final state. (c): Flow surrounding wound boundary for identical initial conditions but varying values of $\pert$. (d): Wound areas throughout closure for different values of $\epsilon_\alpha$. Panels (a-d) plotted with $\Lambda=0.1$, $R_1^0(0)=0$, $R_2^0(0)=0$. (e): Experimental observation of correlation between wound anisotropy and degree of nematic order in the surrounding epithelium. $||\langle q_{...}\rangle ||$ corresponds to average magnitude of $q$ tensor for cells/wounds, averaging over all cells/wounds in first $\sim 10$ minutes (5 frames) of each time-lapse. Points are different wounds, with colour indicating value of $\theta_\mathrm{wound}$ -- illustrated in (f) -- the angle between the tissue axis and the major axis of the wound polygon averaged over initial $\sim 10$ minutes.}
\label{fig:flowAndShape}
\end{figure*}

\paragraph*{\label{sec:order0}Passive bulk tissue with active hole.}
The flow at leading order is radial by symmetry and independent of $\bv{Q}$. Solving for $p^0, \bv{v}^0$ (see SI), we substitute into the KBC to obtain an ODE for the inner mode radius (momentarily reinstating dimensional quantities)
\begin{equation}
\dot{R}_1^0 = -\frac{\gamma R_2^0 (R_2^0+R_1^0) + P R_1^0 (R_2^0)^2}{2\eta [(R_2^0)^2-(R_1^0)^2] + \Gamma (R_1^0 R_2^0)^2 \log \frac{R_2^0}{R_1^0}} \;,
\end{equation}
where $P$ is the difference in external pressures at the inner and outer boundaries, $P=p_2^\mathrm{ext} - p_1^\mathrm{ext}$. We see that, at leading order, closure is driven by the effective surface tension (which includes both passive surface tension and active purse-string contributions) and positive pressure differences ($P>0$) and is slowed by (dissipative) viscosity, friction and negative pressure differences ($P<0$).

\paragraph*{\label{sec:nematic}Nematic Texture and Active Stresses.}
When we switch on activity ($\alpha \ne 0$), the shape anisotropy, $\bv{Q}$ becomes relevant, generating a nematic texture between the inner and outer radii $R_1^0(t), R_2^0(t)$. Minimisation of free energy in eqn.~\eqref{eqn:LdG} results in a non-linear PDE, which we linearise by expanding around the homogeneous ordered state, aligning with the $x$-axis: $Q_{xx}^0 = 1 + q_1, Q_{xy}^0=q_2$, to obtain: $\nabla^2 q_1 - \Lambda^2 q_1 \approx 0\,, \quad  \nabla^2 q_2 \approx 0$, a good approximation provided $|q_1|, |q_2| \ll 1$. 
We take parallel anchoring conditions on $R_1^0$, and $x$-axis alignment conditions on $R_2^0$ (eqn.~SX). We obtain an active stress distribution given by 
\begin{equation}\label{eqn:activeDriving}
\begin{split}
& \nabla \cdot \bv{Q}^0 = - \left[G_2(r) \sin (2 \theta ) + G_4(r) \sin (4 \theta )\right] \uVec_\theta \\ 
  &\,+ \left[G_0(r) + G_2(r) \cos (2 \theta ) + G_4(r) \cos (4 \theta ) \right] \uVec_r \;.
\end{split}
\end{equation}
See SI for the expressions for $G_i(r,\Lambda)$. We point out the presence of $n=0, 2$ and $4$ modes in the driving force appearing in the Stokes equation at $\order (\pert)$, which will drive the free surfaces away from circular. Fig.~\ref{fig:flowAndShape}a illustrates the nematic texture and active driving surrounding the wound free boundary. We also highlight the presence of two $-\frac{1}{2}$ topological defects in the alignment field which, as the wound closes,  move towards its centre. Since the wound itself provides an additional $+1$ topological charge, upon closure, the tissue returns to a topologically neutral `healed' state. An illustration of the defect motion during closure is illustrated in Fig.~\ref{fig:flowAndShape}b.

\paragraph*{\label{sec:order1}Activity drives closure and non-circular wound shapes.}
To compactly express the flow at $\order (\pert)$, it is convenient to use the stream function $\bv{v}^1 = \nabla \times (\psi^1 \uVec_z)$ of the velocity field. Taking the curl of the Stokes equation at $\order (\pert)$, the stream function satisfies a driven, modified biharmonic equation:
\begin{equation}\label{eqn:modifiedBiharmonic}
\nabla^4 \psi^1 - \nabla^2 \psi^1 = F_2(r, \Lambda)\sin 2\theta + F_4(r, \Lambda) \sin 4\theta \;.
\end{equation}
See SI for the expressions for $F_i(r,\Lambda)$. After solving for the $p^1, \bv{v}^1$, the KBCs provide a series of ODEs for the dynamics of shape mode amplitudes, $\dot \xi_i^n,\dot \eta_i^n$, $n=0,2,4$ (see eqn. \ref{eq:Fourier-mode}, eqn.~S36). Crucially, we observe that the circular ($\xi_i^0$) and quadropolar ($\xi_i^{2,4}$) modes are driven, arising from the active forces in the bulk, eqn.~\eqref{eqn:activeDriving}. Numerically integrating these ODEs (Fig. SY) determines the boundary shapes and solves the problem. Figure~\ref{fig:flowAndShape}a shows a snapshot of the wound boundary and surrounding flow.

\paragraph*{Discussion:}
Our first main result is that stresses of the form $\sim \alpha \bv{Q}$ drive the free surfaces away from circular (Fig.~\ref{fig:flowAndShape}c), the degree of anisotropy increasing with $\pert$ and elongating parallel (perpendicular) to the $x$-axis in the contractile, $\pert>0$ (extensile, $\pert<0$) case. This prediction is validated by our experimental observations: plotting the degree of nematic order in the tissue, $\langle || q_\mathrm{tissue} ||\rangle$ (averaging over all cell polygon $||q||$'s in the first five frames $= 10$ minutes of each time-lapse) against the degree of nematic order in the wound polygon, $\langle ||q_\mathrm{wound}|| \rangle$ (averaging over wound polygon $||q||$'s in the first 5 frames) indicates that samples with stronger nematic ordering in the bulk are those with greater anisotropy in the wound shape. This is consistent with our model, which suggests that systems with larger $Q_0$ (and therefore $\epsilon_\alpha$) are driven further from circular and provides evidence for the presence of `active-nematic-like' forces in the bulk pupal wing epithelium. Further, by measuring the angle between the tissue axis and wound's major axis for each sample, $\theta_\mathrm{wound}$  (Fig.~\ref{fig:flowAndShape}e, SI) we find that wounds tend to elongate \textit{parallel} to the tissue axis (we find an average $\theta_\mathrm{wound} /\frac{\pi}{2} = 0.093 \pm 0.193$, consistent with zero). Elongation along the tissue axis is a signature of \textit{contractility} in our model (Fig.~\ref{fig:flowAndShape}c), suggesting the surrounding tissue is under a state of contractile stress.

This is consistent with another prediction of our model, namely that contractility ($\pert>0$) \textit{accelerates} closure whilst extensility ($\pert<0$) \textit{delays} it (Fig.~\ref{fig:flowAndShape}c). This is a consequence of the parallel anchoring conditions on $\bv{Q}$ chosen at the wound free surface; we find that taking normal anchoring at the wound edge interchanges the effects of contractility/extensility on rates of closure (see Fig.~SX).

\paragraph*{Conclusion:}
Informed by experimental observations of re-epithelialisation, we have solved the free-boundary problem of a closing hole in an active nematic fluid. We find that for parallel anchored nematic alignment at the wound hole, contractile active stresses in the bulk tissue (a) accelerate healing and (b) drive elongation of the wound along the axis of nematic order in the far-field. Given one would expect the tissue to use any mechanism at its disposal to accelerate healing, our model suggests the tissue is under a state of contractile active stress -- a prediction that is supported by our measurements of wound anisotropy and alignment parallel to the axis of nematic order in the surrounding tissue.

We have taken the alignment conditions at the boundaries as fixed inputs to the model -- i.e. we do not explain what causes the parallel anchoring in our experiments nor do we allow for dynamics in this nematic order. The precise nature of the mechanism that causes specific cellular alignments at epithelial gaps is the subject of ongoing research \cite{perez-verdugo_anisotropic_2025}. However, given that we observe tangential cellular alignment on a time-scale shorter ($\sim 5 - 10$ minutes) than the time-scale for formation of the actin purse-string (reported as $\sim 15-30$ minutes for wounds in the notum epithelium of \textit{Drosophila} at a similar stage of development \cite{antunes_coordinated_2013}), we believe the release of mechanical tension and recoil following laser ablation is relevant here and will provide the focus for future work.

Although we have focused on re-epithelialisation of wounds in \textit{Drosophila} pupae, this work highlights the relevance of tissue structure and bulk force generation to modelling epithelial gaps more generally. It would be interesting to see how these ideas extend to understanding the mechanics of epithelial closure events in other biological contexts; for example \textit{Ciona} neural tube closure \cite{hashimoto_sequential_2015} or \textit{Drosophila} dorsal closure \cite{de_septate_2022}, where cells tend to align perpendicular to the epithelial gap boundaries instead.

\begin{acknowledgments}
TBL, JT, IVC, RRB, HA would like to thank the Isaac Newton Institute for Mathematical Sciences, Cambridge, for support and hospitality during the programme {\em New statistical physics in living matter}, where part of this work was done. This work was supported by EPSRC grants EP/R014604/1 and EP/T031077/1. HA acknowledges the support of an EPSRC studentship and thanks Luke Neville for helpful discussions and advice regarding calculational details. JT is supported by the Eric and Wendy Schmidt AI in Science Postdoctoral Fellowship. This work was carried out using the computational facilities of the \href{http://www.bristol.ac.uk/acrc/}{Advanced Computing Research Centre, University of Bristol}.
\end{acknowledgments}

%

\newpage
\onecolumngrid
\setcounter{equation}{0}
\setcounter{section}{0}
\setcounter{figure}{0}
\setcounter{table}{0}
\setcounter{page}{1}
\renewcommand{\theequation}{S\arabic{equation}}
\renewcommand{\thefigure}{S\arabic{figure}}
\allowdisplaybreaks[1]

\begin{center}
  \textbf{\large Dynamics of Wound Closure in Living Nematic Epithelia\\ Supplementary Information}\\[.2cm]
  Henry Andralojc,$^{1,*}$ Jake Turley,$^{1,2,3}$ Helen Weavers,$^3$ Paul Martin,$^{3}$ Isaac V. Chenchiah,$^1$ Rachel R. Bennett,$^{1,4,\dagger}$ and Tanniemola B. Liverpool$^{1,4,\ddag}$ \\[.1cm]
  {$^1$\textit{School of Mathematics, University of Bristol, Bristol, BS8 1UG, United Kingdom}\\}
    {$^2$\textit{Mechanobiology Institute, National University of Singapore, Singapore, Singapore}\\}
      {$^3$\textit{School of Biochemistry, University of Bristol, University Walk, Bristol, BS8 1TD, UK}\\}
        {$^4$\textit{Isaac Newton Institute for Mathematical Sciences, 20 Clarkson Rd, Cambridge CB3 0EH, UK}\\}
\end{center}

\section{Experiment Analysis}
\subsection{Cell Shape}
Our analysis of the epithelial cell shapes, elongation and alignments follows that outlined previously in \cite{olenik_fluctuations_2023, turley_deep_2023}. Following the segmentation of epithelial cell boundaries using a bespoke neural-network-based algorithm -- described in \cite{turley_deep_2024} -- each cell in a given frame of the time-lapse was approximated by a polygon. The area, centroid and shape of each polygon, labelled by index $i$, were then quantified by calculating various moments. The first moment defines the cell's area:
\begin{subequations}\label{eqn:sMoments}
\begin{equation}
    A_i = \iint_{i^\text{th} \text{cell}} \dd x \dd y \, ,
\end{equation}

where the integral $\iint_{i^\text{th} \text{cell}}$ is taken over the $i^\text{th}$ cell. The next moment defines the cell's centroid:
\begin{equation}\label{eqn:sCentroid}
    c_i = \frac{1}{A_i}  \iint_{i^\text{th} \text{cell}} 
    \begin{pmatrix}
        x \\ 
        y
    \end{pmatrix} \dd x \dd y \,,
\end{equation}
and the next defines the cell's inertia tensor:
\begin{equation}
    s_i = \frac{1}{A_i^2} \iint_{i^\text{th} \text{cell}} \begin{pmatrix}
        -y^2 & xy \\
        xy & - x^2 
    \end{pmatrix} \dd x \dd y 
\end{equation}
The traceless version of this tensor, defined as:
\begin{equation}\label{eqn:sCellQs}
    q_i = s_i - \frac{\mathbb{1}}{2} \trace[s_i]=
    \begin{pmatrix}
        q_i^{xx} & q_i^{xy} \\
        q_i^{xy} & q_i^{xx}
    \end{pmatrix} = \frac{1}{2 A_i^2} \iint_{i^\text{th} \text{cell}} \begin{pmatrix}
        x^2-y^2 & 2xy \\
        2xy & y^2 - x^2 
    \end{pmatrix} \dd x \dd y 
\end{equation}
\end{subequations}
was used to quantify each cell polygon's alignment relative to the tissue as a whole; a large, positive $q^{xx}$ corresponds to a cell mostly aligned along the Cartesian $x$-axis (which in this case roughly aligns with the PD-axis of the developing wing), large negative values of $q^{xx}$ correspond to cells largely aligned along the $y$-axis (i.e. the wing AP-axis). The off-diagonal components of $q$ quantify the alignment along the lines $y = \pm x$. See Figure~\ref{fig:sQTensors}.

The anisotropy in shape of the $i^\text{th}$ cell can equivalently be quantified in terms of a `magnitude' and angle:
\begin{equation}\label{eqn:sQMagAngle}
    ||q_i||^2 = \frac{1}{2} \trace[q_i^2] = (q_i^{xx})^2 + (q_i^{xy})^2 \, , \quad \phi_i = \frac{1}{2} \arctan \left(\frac{q^{xy}_i}{q^{xx}_i} \right)
\end{equation}

Large $||q_i||$ then corresponds to cells with large aspect-ratio (highly elongated cells) and $\phi_i$ gives the angle of the major axis of the cell with respect to the $x$-axis. Figure 1a in the main text displays each of the segmented cells in an unwounded sample, with fill color given by $||q_i||$ in the left half and $\phi_i$ in the right half. Averages displayed in Fig. 1b are taken over all cells visible in the frame.

\subsection{Wound Shape and Alignment between Tissue and Wound Anisotropy}
Wound outlines were identified using a semi-automated process, outlined fully in \cite{turley_deep_2024, turley_deep_2023}. To summarise, the ImageJ plugin `Trainable Weka Segmentation' (a supervised-learning machine learning algorithm) was trained to find areas of the images that are tissue or non-tissue. Non-tissue included both the wound (if present) and any parts of the frame where the tissue was folded or outside the imaging region. Tissue/non-tissue binary masks were then hand-edited to remove errors, primarily due to debris surrounding the wound site. The wound outlines in each frame were then segmented and approximated by polygons, and moments calculated using equations~\eqref{eqn:sMoments}.

In our measurements of wound anisotropy, see Figure 2e, we average over the magnitude of the wound's $q$ tensor, $q_\mathrm{wound}$ for the first 5 frames (corresponding to $\sim 10$ minutes) of the time-lapse. That is,
\begin{equation}
    \langle ||q_\mathrm{wound}|| \rangle = \frac{1}{5} \sum_{\text{wounds in first 5 frames}} ||q_\mathrm{wound}||\,.
\end{equation}

The degree of elongation in the tissue in Fig. 2e is calculated similarly, averaging over $q_i$'s for all cells observed in the first five frames of each time-lapse:
\begin{equation}
    \langle ||q_\mathrm{tissue}|| \rangle = \frac{1}{N} \sum_{i\in \text{cells in first 5 frames}}^N ||q_i|| \,,
\end{equation}
where $N$ is the total number of cell polygons segmented in the first 5 frames. The polygons in one frame aren't connected in any way with polygons in later frames (i.e. there are as many polygons corresponding to the same cell as there are frames which feature that cell). Therefore this sum includes all polygons segmented in first frame $+$ all polygons segmented in second frame etc... The misalignment between the axis of the tissue and the major axis of the wound's shape is then calculated in the following way. The angle made by the wound's major axis to the $x$-axis was averaged over the first five frames:
\begin{equation}
    \phi_\mathrm{wound} = \frac{1}{5} \sum_{j\in \text{wounds in first 5 frames}} \phi_j \, ,
\end{equation}
and the angles made by cell's major axes to the $x$-axis was averaged:
\begin{equation}
    \phi_\mathrm{tissue} = \frac{1}{N} \sum_{i \in \text{cells in first 5 frames}} \phi_i \,.
\end{equation}

The misalignment between these two axes, $\theta_\mathrm{w} = \phi_\mathrm{wound} - \phi_\mathrm{tissue}$ is displayed in Figure 2e using colour to indicate magnitude. Each point in Fig. 2e corresponds to a different pupa.

\subsection{Rotation into `Wound' Basis}
\begin{figure}
    \centering
    \includegraphics[width=0.5\linewidth]{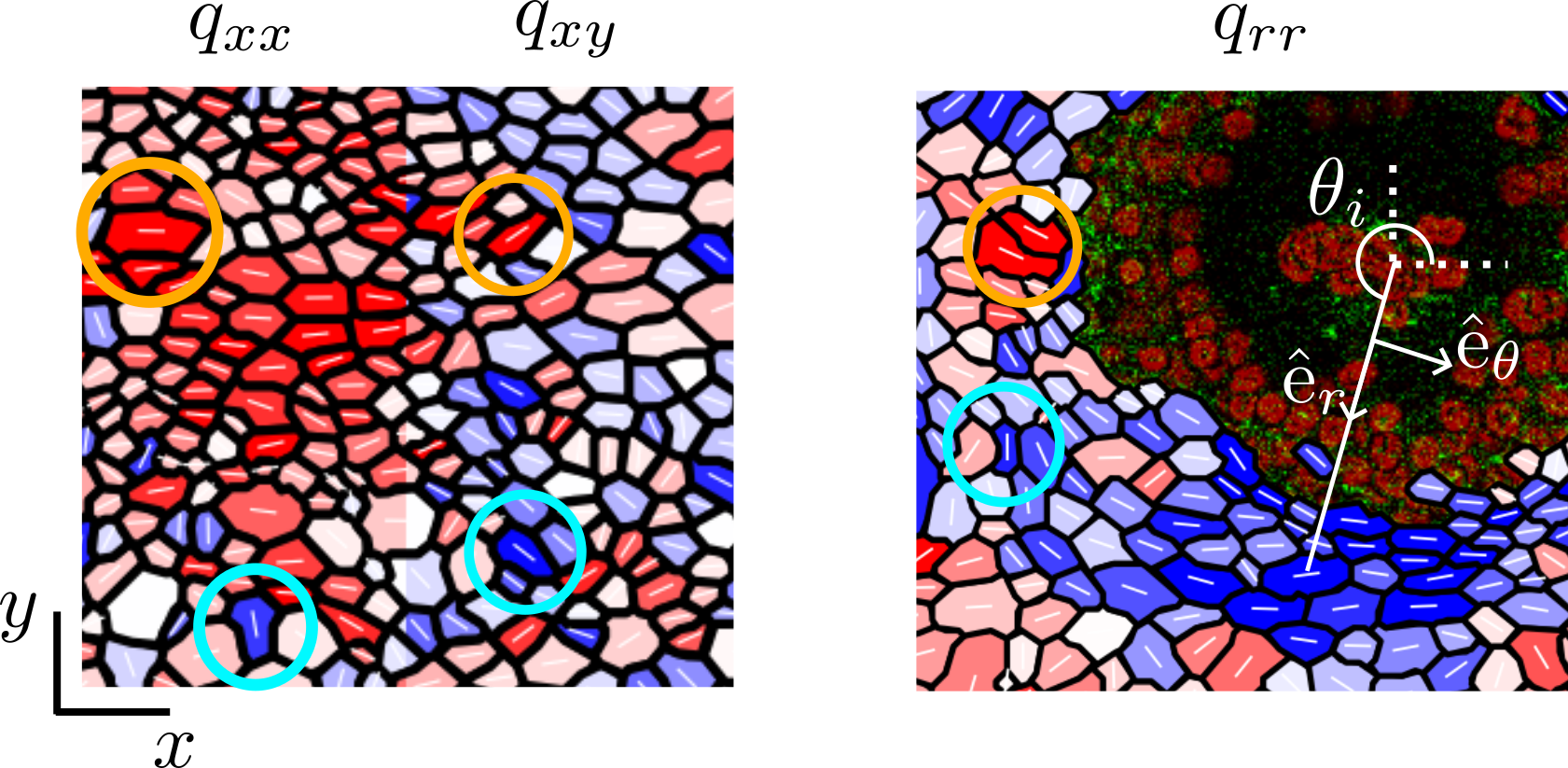}
    \caption{\raggedright Shape and orientation of each cell is described by traceless tensor $q_i$. (a) Examples of segmented cells with colour code indicating magnitude of $q^{xx}$ ($q^{xy}$) in the left (right) panel. Orange and cyan circles in left panel indicate cells that align strongly along the $x$ and $y$ axes (and so have large positive/negative values of $q^{xx}$) respectively. Orange and cyan circled cells in the right panel align strongly along the lines $y = x$ and $y=-x$ respectively. (b) Segmented cells surrounding a wound, colour coded by magnitude of $q^{rr}$. The orange circle highlights a cell elongated along the $\uVec_r$ polar unit vector, and the cyan circle highlights a cell elongated along the $\uVec_\theta$ polar unit vector.}
    \label{fig:sQTensors}
\end{figure}

In order to quantify the effect of the wound on cellular alignment, it is useful to perform a change of basis by rotating each $q_i$ in a polar basis centred on the wound's centroid, defined analogously to eqn.~\ref{eqn:sCentroid} (instead integrating over the polygon approximating the wound). The components of this tensor then give information about the alignment along the radial and tangential unit vectors in a basis centred on the wound, see Fig.~\ref{fig:sQTensors}. Denoting the angle subtended from the $x$-axis (centred on the wound's centroid) to the $i^\text{th}$ cell as $\theta_i$, we transform each $q_i$ into a local polar basis:
\begin{equation}
    q_i \to R(\theta_i) q_i R^\mathrm{T}(\theta_i) = 
    \begin{pmatrix}
        \cos \theta_i & \sin \theta_i \\
        - \sin \theta_i & \cos \theta_i 
    \end{pmatrix}
    \begin{pmatrix}
        q_i^{xx} & q_i^{xy} \\
        q_i^{xy} & - q_i^{xx}
    \end{pmatrix}\begin{pmatrix}
        \cos \theta_i & -\sin \theta_i \\
        \sin \theta_i & \cos \theta_i 
    \end{pmatrix} = 
    \begin{pmatrix}
        q_i^{rr} & q_i^{r\theta} \\
        q_i^{r\theta} & - q_i^{rr}
    \end{pmatrix}
\end{equation}
where the rotation matrix, $R(\theta) = \big(\begin{smallmatrix}
  \cos \theta  & \sin \theta \\
  -\sin \theta  & \cos \theta 
\end{smallmatrix}\big)$. Large positive values of $q^{rr}$ correspond to cells aligned predominantly along the $\uVec_r$ vector, whilst large negative values correspond to cells aligned along $\uVec_\theta$. Analogously to eqn.~\eqref{eqn:sQMagAngle}, it is useful to instead look at the `magnitude' and an angle:
\begin{equation}
    ||q_i||^2 = \frac{1}{2} \trace [q_i^2] \,, \quad \tilde{\phi}_i = \frac{1}{2} \arctan\left(\frac{q_i^{r\theta}}{q_i^{rr}}\right)
\end{equation}
where the angle $\tilde{\phi}_i$ is now measured with respect to the $\uVec_r$ vector direction. Figure 1c of the main text actually plots $\phi^\prime_i = \tilde{\phi}_i + \pi/2$ (and so gives the angle of cellular alignment measured with respect to $\uVec_\theta$) which is then wrapped onto $\phi^\prime_i \in [-\pi/2, \pi/2]$ for visualisation.

\section{Model Setup}
\subsection{Equations of Motion and Boundary Conditions}
We model the flow of tissue surrounding the wound using the equations of active nematohydrodynamics at vanishing Reynolds number:
\begin{subequations}
\begin{gather}
0 = \nabla \cdot \bv{v} \;, \\
0 = -\Gamma \bv{v} + \nabla \cdot \boldsymbol{\sigma}  \;, \\
0 = \funcdiff{\mathcal{F}_\mathrm{LdG}}{\bv{Q}} \;. \label{eqn:sRelaxQ}
\end{gather}
\label{eqn:sEquationsOfMotion}
\end{subequations}
taking
\begin{equation}
    \sigma_{ij} = -p \delta_{ij} + \eta (\partial_i v_j + \partial_j v_i) + \alpha Q_{ij} \;,
\end{equation}
where $\bv{v}, p$ are the fluid velocity and pressure respectively and $\bv{Q}$ is the tensorial nematic order parameter describing local alignment of cells in the tissue. $\eta$ is the bulk viscosity and $\Gamma$ is the frictional coefficient, included to model the resistance to flow due to the presence of a substratum below the epithelium. We take Landau-de Gennes free energy:
\begin{equation}\label{eqn:sLdG}
    \mathcal{F}_{\mathrm{LdG}} = \int \dd ^2 x \left[-\frac{A}{2} ||\bv{Q}||^2 + \frac{B}{4} ||\bv{Q}||^4 + \frac{K}{2} ||\nabla \bv{Q}||^2 \right]\;,
\end{equation}
and $||\bv{Q}||^2 = Q_{ij}Q_{ij}, \, ||\nabla \bv{Q}||^2 = \partial_k Q_{ij} \partial_k Q_{ij}$ (summation over repeated indices implied). We consider the relaxation in the cellular shape to be much faster than the flow (cell motion) time scale, allowing us to approximate the  nematic texture, $\bv{Q}$ to be always in a local equilibrium (dependent on boundary conditions). Functional minimisation of free energy in equation~\eqref{eqn:sLdG} with respect to $Q_{ij}$ results in non-linear PDE for $\bv{Q}$:
\begin{equation}
    0 = -A Q_{ij} + B Q_{ij} ||\bv{Q}||^2 - K \nabla^2 Q_{ij}\;. 
\end{equation}

The dynamic boundary conditions on the flow are ($i=1,2$):
\begin{equation}\label{eqn:sDBCs}
\left[\left(p^\mathrm{ext}_i - p\right) \bv{n} + (2\eta \bv{D} + \alpha \bv{Q}) \cdot \bv{n} \right] \big |_{R_i} = \gamma \kappa_i \bv{n} \;,
\end{equation}
where $D_{ij} = \frac{1}{2} \left(\partial_i v_j + \partial_j v_i\right)$ is the strain rate and $p_i^{\mathrm{ext}}$ is the pressure in the external fluid pressure in contact with the $i^\text{th}$ free surface (see Fig.~1X). The effective surface tension $\gamma= \tilde{\gamma} + \tau $ includes both passive ($\tilde{\gamma}$) surface tension and an active component ($\tau$) due to a possible contractile purse string at the wound boundary.

The kinematic boundary condition governs the time evolution of the free boundaries:
\begin{equation}
	D_t(r-R_i(\theta,t)) = 0 \quad \Rightarrow \quad v_r |_{R_i} = \dot{R}_i + \frac{v_\theta |_{R_i}}{R_i} R^\prime_i \;.
\end{equation}

This boundary condition states that an indicator function $\mathcal{I}=r-R_i(\theta,t)$, which is non-zero everywhere except on the free surface, must be convected by the flow -- that is, $\mathcal{I}$ must remain zero on the boundary for all times. The material derivative $D_t = \partial_t + \bv{v} \cdot \nabla$ is required here to include the effect of the flow in an Eulerian frame, see e.g. \cite{batchelor_introduction_2000}.

We take parallel anchoring conditions on $\bv{Q}$ at the inner boundary:
\begin{subequations}
\begin{equation}
    \begin{split}
    & Q_{xx} |_{R_1} = S_1 \frac{\cos 2\theta \left[(R_1^\prime)^2  - R_1^2\right] - 2 R_1^\prime R_1 \sin 2 \theta}{R_1^2+(R_1^\prime)^2} \;,\\
    & Q_{xy} \big |_{R_1} = S_1 \frac{\sin 2\theta \left[(R_1^\prime)^2  - R_1^2\right] + 2 R_1^\prime R_1 \cos 2 \theta}{R_1^2+(R_1^\prime)^2} \;,
    \end{split}
\end{equation}
\end{subequations}
and $x$-axis alignment at the outer boundary:
\begin{equation}
    Q_{xx} \big |_{R_2} = S_2 \;,\\
    Q_{xy} \big |_{R_2} = 0 \;.
\end{equation}

$S_i$ denotes the scalar nematic order parameter on the $i^\text{th}$ boundary, $S_i = 0 \Rightarrow$ isotropic cells.

Before proceeding with non-dimensionalisation and expanding in activity, we take a moment to clarify some of the choices in our model and relation to our experimental measurements. 

\subsection{Inner and Outer Boundary}
As discussed in the main text, we model the epithelium as an incompressible active nematic fluid. The assumption of incompressibility is the result of  our observation of a decrease in the cellular division rate in wounded tissue immediately after wounding~\cite{turley_deep_2024}. In order to have a well-posed mathematical problem which reproduces closure of the inner boundary, we specify the boundary conditions in all our fields at the wound `inner' boundary ($r=R_1(\theta,t)$) and also an `outer' boundary ($r=R_2(\theta,t)$) at finite distance. We therefore track the evolution of two free-surfaces. The balance fluxes due to incompressibility at the inner/outer boundaries (for axisymmetric situations) implies that $ R_1 v_r |_{R_1} \overset{\text{KBC}}{=} R_2 v_r |_{R_2}$ so provided $R_2 \gg R_1 \, \Rightarrow \;  \dot R_2 \ll \dot R_1$. As we note below, for realistic situations, the results do not depend appreciably on the conditions at outer boundary. 

\subsection{Effective Surface Tension at Wound Boundary}
Although our wound healing experiments do not stain for actin or myosin -- and so we cannot claim a purse string mechanism is driving closure of our wounds -- contractile actomyosin cables have been observed around wounds in \textit{Drosophila} at similar stages of development, for example \cite{antunes_coordinated_2013} where pupae were wounded in the notum epithelium at 13 hours after puparium formation (APF) (vs. 18hAPF in our experiments).

Following \cite{cochet-escartin_border_2014, almeida_mathematical_2011}, we assume that the contractile actomyosin `purse string' would provide a constant force that goes as the curvature $\sim \kappa$ -- i.e. acts like a surface tension. The curvature is relevant here because the actomyosin cable acts as a line under tension directed along the direction tangential to the wound boundary. When the wound boundary has some curvature, the net force at any element of the wound boundary $\bv{F} \sim \tau \delta s \dd \bv{t} /\dd s \sim \tau \kappa \bv{n} \delta s$, where $\bv{t}$ is the tangent vector to the boundary parametrised by arc-length $s$ and $\delta s$ is a small element of the boundary. More sophisticated models which incorporate heterogeneity in the force provided by the purse string have been considered by \textcite{almeida_modeling_2012, roldan_computational_2019}, for example. Since our focus here is to understand the effect of bulk activity (the $\alpha \bv{Q}$ term), we take this contribution as a homogeneous constant for simplicity.

We also comment that the purse string is not necessary for closure in our model. Passive surface tension ($\tilde{\gamma}$), differences between external pressures ($P=p_\mathrm{ext}^2 - p_\mathrm{ext}^1$) and bulk activity (with an appropriate sign of $\alpha$) would all still drive closure. The purse string contribution simply accelerates closure by increasing the effective surface tension.

\subsection{Order parameter $\bv{Q}$ and `microscopic' $q_i$}
We also wish to clear any possible confusion between the shapeless shape tensor $q_i$ -- which we measure for each of the segmented cells in our experiments -- and the order parameter $\bv{Q}$ in our model. One might make the connection between the two by viewing $\bv{Q}$ as an object describing the average local nematic alignment in the tissue, that is a coarse-grained/smoothed version of a `spikey' microscopic field $\sum_i q_i \delta(\bv{r}-\bv{c}_i)$:
\begin{equation}
    \rho(\bv{r}) \bv{Q}(\bv{r}) = \sum_i \iint \dd^2 \bv{r}^\prime K(\bv{r}^\prime - \bv{r}) q_i \delta(\bv{r}^\prime - \bv{c}_i) \, ,
\end{equation}
where the coarse-grained density
\begin{equation}
    \rho(\bv{r}) = \sum_i \iint \dd^2 \bv{r}^\prime K(\bv{r}^\prime - \bv{r}) \delta(\bv{r}^\prime - \bv{c_i}) \, ,
\end{equation}
and $K( ... )$ is some finite-ranged averaging kernel and $\bv{c}_i$ are the centroids of each cell (see eqn.~\eqref{eqn:sCentroid}).

\subsection{Non-Dimensionalisation}
We non-dimensionalise by introducing length, time and pressure scales, and units for $\bv{Q}$: $\bv{r} \to L \bv{r},\, t \to T t,\, p \to \Pi p, \, \bv{Q} \to Q_0 \bv{Q}$, choosing:
\begin{equation}
    L =  \sqrt{\frac{\eta}{\Gamma}}, \quad T = \frac{\eta}{\Pi}, \quad \Pi = \frac{\gamma}{L}, \quad Q_0 = \sqrt{\frac{A}{2B}} \;.
\end{equation}

The non-dimensional equations of motion,
\begin{subequations}
\begin{gather}
0 = \nabla \cdot \bv{v}\;, \\
0 = - \nabla p + \nabla^2 \bv{v} - \bv{v} + \pert (\nabla \cdot \bv{Q})\;, \\
0 = \Lambda^2 Q_{ij} \left[(Q_{xx}^2 + Q_{xy}^2) - 1\right] - 2 \nabla^2 Q_{ij}\;, \label{eqn:sNonDimQEqn}
\end{gather}
\end{subequations}
now contain two dimensionless parameters, $\pert = \alpha Q_0/\Pi$ and $\Lambda = L/\ell_Q$. $\pert$ determines the strength of bulk active stresses relative to other passive stresses and $\Lambda$ is the ratio of flow ($L^2 = \eta/\Gamma$) to nematic length scales ($\ell_Q^2 = K/2A$) and governs the persistence of active stresses into the bulk. $1/\Lambda$ gives a proxy for the nematic length scale in the system.

\subsection{Perturbation Parameter}
Our experimental wounds are approximately circular throughout closure, so we choose to expand around circular free boundaries, using the non-dimensionalised activity $\pert = \alpha Q_0 / \Pi$ as the perturbation parameter:
\begin{subequations}\label{eqn:sExpansion}
\begin{gather}
R_i(\theta,t) = R_i^0(t) + \pert R_i^1(\theta,t) \\
\bv{v} = \bv{v}^0 + \pert \bv{v}^1 + ... \\
p = p^0 + \pert p^1 + ... \\
\bv{Q} = \bv{Q}^0 + \pert \bv{Q}^1 + ...
\end{gather}    
\end{subequations}

In this way, the leading order problem is that of a passive droplet in an axisymmetric annulus of with inner and outer radii $R_1^0(t)$, $R_2^0(t)$. Further, the first order contributions to the flow only `see' leading order contributions from the nematic driving. This can be seen by substituting the expansion into the non-dimensionalised Stokes equation:
\begin{equation*}
\begin{split}
0 = - \nabla p^0 + \nabla^2 \bv{v}^ 0 - \bv{v}^0 + \pert \left[-\nabla p^1 + \nabla^2 \bv{v}^1 - \bv{v}^1 + \nabla \cdot \bv{Q}^0\right] + \order(\pert^2).
\end{split}
\end{equation*} 

This expansion approximation should provide accurate qualitative predictions provided the non-circularity $\pert R_i^1 / R_i^0 $ remains small, which since we consider circular initial conditions, will be satisfied for short times immediately following wounding. This is appropriate for a continuum model for re-epithelialisation, as we expect that as the wound size approaches the scale of individual cells, other mechanisms will become important.

\subsection{Model Wound Area and Anisotropy}
We will find that the $\alpha \bv{Q}$ term only contributes $n=0,2,4$ modes to the wound shape. Therefore, we expand the wound radius in a Fourier series:

\begin{equation}
    R_1(\theta,t) = R_1^0(t) + \pert \sum_{k=0,2,4} \xi_1^k(t) \cos k\theta + \eta_1^k(t) \sin k\theta
\end{equation}

We can then express quantities like the area and $q$ tensor for the shape enclosed by the inner free surface using equations~\eqref{eqn:sMoments}. The area enclosed by the inner free surface -- i.e. wound area -- $A_\mathrm{w}$, in terms of shape mode amplitudes:
\begin{equation}
\begin{split}
    A_\mathrm{w} &= \iint_\mathrm{wound} \dd x \dd y = \int_0^{2\pi} \dd \theta \int_0^{R_1(\theta,t)} r \dd r = \frac{1}{2} \int_0^{2\pi} [R_1(\theta,t)]^2 \dd \theta \\
    & =\pi \left[(R_1^0)^2 + 2 \pert \xi_1^0 R_1^0 + \pert^2 \left[2(\xi_1^0)^2 + (\xi_1^2)^2 + (\xi_1^4)^2 + (\eta_1^2)^2 + (\eta_1^4)^2 \right] \right] \, ,
\end{split}
\end{equation}
and the $q$ tensor for the shape enclosed by the inner free surface:
\begin{subequations}
\begin{equation}
\begin{split}
    q^{xx}_\mathrm{w} & = \frac{1}{2 A_\mathrm{w}^2} \iint_\mathrm{wound} (x^2-y^2) \dd x \dd y =  \frac{1}{2A_\mathrm{w}^2} \int_0^{2\pi} \dd \theta \int_0^{R_1(\theta,t)} r^3 \cos 2\theta \dd r  = \frac{1}{8A_\mathrm{w}^2} \int_0^{2\pi} \dd \theta \, \cos 2\theta [R_1(\theta,t)]^4 \\ 
    & = \frac{\xi_1^2}{2\pi R_1^0} \pert + \order(\pert^2),
\end{split}
\end{equation}

\begin{equation}
\begin{split}
    q^{xy}_\mathrm{w} & = \frac{1}{A_\mathrm{w}^2} \iint_\mathrm{wound} xy \, \dd x \dd y =  \frac{1}{2A_\mathrm{w}^2} \int_0^{2\pi} \dd \theta \int_0^{R_1(\theta,t)} r^3 \sin 2\theta \dd r  = \frac{1}{8A_\mathrm{w}^2} \int_0^{2\pi} \dd \theta \, \sin 2\theta [R_1(\theta,t)]^4 \\ 
    & = \frac{\eta_1^2}{2\pi R_1^0} \pert + \order(\pert^2) \, ,
\end{split}
\end{equation}
\end{subequations}
from which we can calculate the magnitude, $||q_\mathrm{w}||$ which gives a measure for anisotropy of the wound shape:
\begin{subequations}\label{eqn:sModelWoundQAngle}
\begin{equation}
    ||q_\mathrm{w}|| = \sqrt{(q^{xx}_\mathrm{w})^2 + (q^{xy}_\mathrm{w})^2} = \frac{\sqrt{(\xi_1^2)^2 + (\eta_1^2)^2}}{2\pi R_1^0} |\pert| + O(\pert^2)
\end{equation}

This demonstrates the $n=2$ modes are -- at leading order -- responsible for anisotropy in shape. The angle of the major axis of the wound (measured relative to $x$) is:
\begin{equation}
    \phi_\mathrm{wound} = \frac{1}{2} \arctan ( q^{xy}_\mathrm{w}/q^{xx}_\mathrm{w}) = \frac{1}{2} \arctan \left(\frac{\eta_1^2}{\mathrm{sgn}[\pert] \xi_1^2}\right) - \frac{3((\eta_1^2)^2 \eta_1^4 - \eta_1^4 (\xi_1^2)^2 + 2 \eta_1^2 \xi_1^2 \xi_1^4)}{4 R_1^0[(\eta_1^2)^2 + (\xi_1^2)^2]} \pert + \order(\pert^2)
\end{equation}
\end{subequations}

The sign of $\phi_\mathrm{wound}$ is therefore -- to leading order -- controlled by the sign of $\pert$ and $\eta_1^2$ and the `closeness' to the $x$ and $y$ axes by the sign of $\mathrm{sgn}[\pert]\xi_1^2$. This is because:
\begin{equation}
    \begin{cases}
        |\phi_\mathrm{wound}| < \pi/4 & \text{if } \mathrm{sgn}[\pert]\xi_1^2 >0 \\
        |\phi_\mathrm{wound}| > \pi/4 & \text{if } \mathrm{sgn}[\pert] \xi_1^2 <0
    \end{cases}
\end{equation}

Therefore, (in the contractile case $\mathrm{sgn}[\pert]>0$) when $\xi_1^2>0$, the wound axis is more closely aligned with the $x$-axis than the $y$-axis and when $\xi_1^2<0$, the wound axis is more closely aligned with the $y$ than the $x$-axis. This demonstrates we can make qualitative statements about the wound shape by looking at the signs of the shape mode amplitudes $\xi_1^k(t), \eta_1^k(t)$ and the sign of the active parameter $\pert$. These mode amplitudes will be the final result of our calculation.

\section{Calculational Details}\label{sec:sCalculation}
\subsection{Boundary Conditions at $\order (1)$}
Substituting expansions~\eqref{eqn:sExpansion}, we obtain boundary conditions for each order in $\pert$. At $\order (1)$, the normal and tangential components of the dynamic boundary conditions on each boundary read:
\begin{equation}
\begin{split}
-\frac{\mu_i}{R^0_i} = \left[p_i^\text{ext} - p^0 + 2 \partial_r v_r^0\right] \, \big |_{R_i^0} \;, \\
0 = \left[\frac{\partial_\theta v_r^0}{R_i^0} + \partial_r v_\theta^0 - \frac{v_\theta^0}{R_i^0}\right] \bigg |_{R_i^0} \;,
\end{split}
\end{equation}
where $\mu_i$ is an index $\mu_1 = -1$ for the inner boundary, $\mu_2 = +1$ for the outer boundary. This is included to ensure the relative directions of surface tension and the outward normals are correctly accounted for -- on the inner boundary, surface tension points in the \textit{same} direction as the outward normal, whereas surface tension points in the \textit{opposite} direction to the outward normal on the outer boundary. The kinematic boundary condition at $\order(1)$ reads
\begin{equation}\label{eqn:sKBC0}
v_r^0 \big |_{R_i^0} = \dot{R}_i^0 \;.
\end{equation}

The boundary conditions on the nematic texture at $\order (1)$ are
\begin{subequations} \label{eqn:sQBCs}
\begin{gather}
Q_{xx}^0 \big |_{R_1^0} = - \beta_1 \cos 2\theta \;, \\
Q_{xy}^0 \big |_{R_1^0} = - \beta_1 \sin 2\theta \;, \\
Q_{xx}^0 \big |_{R_2^0} = \beta_2\; , \\
Q_{xy}^0 \big |_{R_2^0} = 0 \;,
\end{gather}
\end{subequations}
where $\beta_i = S_i/(2Q_0)$ is the rescaled nematic scalar order at the $i^\text{th}$ boundary. This can, in principle, be estimated from our experimental data by averaging $||q||$ for cells close to the wound/in healthy tissue. The forms for the boundary conditions at $\order(\pert)$ are more complicated, as they require one to a) replace each dynamical field with its expansion in $\pert$ and b) Taylor expand each term in the first argument around radius $R_i^0$. That is, $v_r \big|_{R_i} = \left[v_r^0 + \pert v_r^1 + ...\right] \big |_{R_i} = \left[v_r^0 + \pert (R_i^1 \partial_r v_r^0 + v_r^1) + ... \right] \big |_{R_i^0}$. Therefore, we momentarily postpone our statement of the boundary conditions on the flow at $\order(\pert)$.

\subsection{Nematic Texture at $\order (1)$}
Equation~\eqref{eqn:sNonDimQEqn} is a non-linear PDE, which we linearise by expanding around the homogeneous ordered state aligning along the $x$-axis: 
\begin{equation*}
\bv{Q}^0 = \begin{pmatrix}
	1 + q_1 & q_2 \\
	q_2 & - 1- q_1
\end{pmatrix}.
\end{equation*}

Substituting into equation~\eqref{eqn:sNonDimQEqn} and ignoring terms $\mathcal{O}(q^2, p^2)$, we obtain two linear PDEs for $q_1,q_2$:
\begin{gather}\label{eqn:sLinearisedQEqns}
 \Rightarrow \quad \nabla^2 q_1 - \Lambda^2 q_1 \approx 0, \quad  \nabla^2 q_2 \approx 0,
\end{gather}
which will provide a good approximation to the nematic field provided $|q_1|, |q_2| \ll 1$. Solutions to equations~\eqref{eqn:sLinearisedQEqns} are sums of modified Bessel functions and powers of $r$ respectively. Substituting into the boundary conditions~\eqref{eqn:sQBCs}, and solving the resulting linear system for unknown constants, we find:
\begin{subequations}
\begin{gather}
q_1(r,\theta,t) = e_0 K_0(\Lambda r) + f_0 I_0(\Lambda r) + \cos 2\theta \left[e_2 K_2(\Lambda r) + g_2 I_2(\Lambda r)\right], \\
q_2(r,\theta,t) = \sin 2\theta \left[\frac{b_2}{r^2} + d_2 r^2 \right],
\end{gather}
\end{subequations}
where $e_0, f_0, b_2, d_2, e_2, g_2$ are all known, but complicated, functions of $R_1, R_2, \Lambda$:
\begin{subequations}
\begin{equation}
e_0 = \frac{(1-\beta_2) I_0\left(\Lambda  R_1^0\right)-I_0\left(\Lambda  R_2^0\right)}{I_0\left(\Lambda  R_2^0\right) K_0\left(\Lambda 
   R_1^0\right)-I_0\left(\Lambda  R_1^0\right) K_0\left(\Lambda  R_2^0\right)},
\end{equation}
\begin{equation}
    f_0 = -\frac{(1-\beta_2) K_0\left(\Lambda  R_1^0\right)-K_0\left(\Lambda 
   R_2^0\right)}{I_0\left(\Lambda  R_2^0\right) K_0\left(\Lambda  R_1^0\right)-I_0\left(\Lambda  R_1^0\right) K_0\left(\Lambda  R_2^0\right)},
\end{equation}
\begin{equation}
   b_2 = \frac{\beta _1
   \left(R_1^0\right)^2 \left(R_2^0\right)^4}{\left(R_1^0\right)^4-\left(R_2^0\right)^4},
\end{equation}
\begin{equation}
d_2 = -\frac{\beta _1
   \left(R_1^0\right)^2}{\left(R_1^0\right)^4-\left(R_2^0\right)^4},
\end{equation}
\begin{equation}
    e_2 =  -\frac{\beta _1 I_2\left(\Lambda  R_2^0\right)}{I_2\left(\Lambda  R_2^0\right) K_2\left(\Lambda 
   R_1^0\right)-I_2\left(\Lambda  R_1^0\right) K_2\left(\Lambda  R_2^0\right)},
\end{equation}
\begin{equation}
g_2 =  \frac{\beta _1 K_2\left(\Lambda  R_2^0\right)}{I_2\left(\Lambda  R_2^0\right)
   K_2\left(\Lambda  R_1^0\right)-I_2\left(\Lambda R_1^0\right) K_2\left(\Lambda  R_2^0\right)} \;.
\end{equation}
\end{subequations}

The form of the driving term appearing in the Stokes eqn. at $\order(\pert)$ is then determined by transforming components of $\bv{Q}^0$ from the Cartesian to polar bases, and taking the divergence in polar coordinates. The symmetry of the tensorial nematic order parameter means that components of $\bv{Q}^0$ in the Cartesian basis are transformed to the polar basis simply as:
\begin{equation}
\begin{pmatrix}
	Q_{rr} & Q_{r\theta} \\
	Q_{r\theta} & - Q_{rr}
\end{pmatrix} = 
\begin{pmatrix}
	\cos 2\theta & \sin 2\theta \\
	-\sin 2\theta & \cos 2\theta
\end{pmatrix}
\begin{pmatrix}
	Q_{xx} & Q_{xy} \\
	Q_{xy} & - Q_{xx}
\end{pmatrix} \;,
\end{equation}
where $\theta$ is the polar coordinate angle. The full form of the driving at $\order(\pert)$ is then:
\begin{equation}
\begin{split}
& \nabla \cdot \bv{Q}^0 = - \left[G_2(r) \sin (2 \theta ) + G_4(r) \sin (4 \theta )\right] \uVec_\theta \\ 
  &\,+ \left[G_0(r) + G_2(r) \cos (2 \theta ) + G_4(r) \cos (4 \theta ) \right] \uVec_r \;,
\end{split}
\end{equation}
where
\begin{subequations}
\begin{equation}
G_0(r) = +2 d_2 r -\frac{1}{2} e_2 \Lambda  K_1(r \Lambda )+\frac{1}{2} g_2 \Lambda  I_1(r \Lambda ),
\end{equation}
\begin{equation}
G_2(r) = -\Lambda  e_0 K_1(r \Lambda ) + \Lambda f_0 I_1(r \Lambda ),
\end{equation}
\begin{equation}
G_4(r) = \frac{2 b_2}{r^3}-\frac{1}{2} e_2 \Lambda  K_3(r \Lambda )+\frac{1}{2} g_2 \Lambda  I_3(r \Lambda ).
\end{equation}
\end{subequations}

And for reasons that will become apparent later, we also have:
\begin{equation}\label{eqn:sCurlDivQ0}
\begin{split}
\left[\curl \left(\nabla \cdot \bv{Q}^0\right)\right]_z &= -\Lambda ^2 \sin 2 \theta \left[f_0 I_2(\Lambda r)+e_0 K_2(\Lambda r)\right] \\
& - \sin 4 \theta \left[\Lambda ^2 \left[e_2 K_4(\Lambda r)+g_2 I_4(\Lambda r)\right]-\frac{12 b_2}{r^4}\right].
\end{split}
\end{equation}

\subsection{Flow at $\order (1)$}
The velocity field at can be determined straightforwardly in the passive problem by invoking a symmetry argument and integrating the incompressibility condition. However, once axisymmetry no longer applies, we will not be able to assume radial flow. Therefore, we spend a moment considering solutions to the Stokes equation of the form:
\begin{equation}\label{eqn:sOrder1Stokes}
\nabla p = \nabla^2 \bv{v} - \bv{v} \;.
\end{equation}

The solutions to the Stokes equation at $\mathcal{O} (\pert)$ will then share the same complementary function, with an additional particular integral arising from nematic driving. We momentarily drop superscripts ${}^0$ for clarity. It will be convenient to use the stream function, defined as $\bv{v} = \curl (\psi \uVec_z)$, where $\psi$ is the stream function and $\uVec_z$ is the unit vector pointing out of the plane. Taking the curl of equation~\eqref{eqn:sOrder1Stokes} and replacing $\curl \bv{v} = \omega_z \uVec_z = - \nabla^2 \psi \uVec_z$ (where $\omega_z$ is the vorticity component in the $z$-direction), we obtain a modified biharmonic equation:
\begin{equation}\label{eqn:sModifiedBiharmonic}
0 = \nabla^4 \psi - \nabla^2 \psi \;.
\end{equation}

Solutions to equation~\eqref{eqn:sModifiedBiharmonic} may be categorised into four apparently distinct families, those for which:
\begin{enumerate}
	\item $\nabla^2 \psi_1 = 0$,
	\item $\nabla^2 \psi_2 - \psi_2 = 0$,
	\item $\nabla^2 \psi_3 = \phi_1$ where $\nabla^2 \phi_1 - \phi_1 = 0$,
	\item $\nabla^2 \psi_4 - \psi_4 = \phi_2$ where $\nabla^2 \phi_2 = 0$.
\end{enumerate}

Let the space of solutions to the Laplace and Helmholtz equations be denoted as $L$ and $H$ respectively. Then clearly $\psi_1 \in L$, $\psi_2 \in H$. $\psi_3$ has complementary functions $\psi_3^\mathrm{CF} \in L$ and particular integral $\psi_3^\mathrm{PI} = \phi_1 \in H$. Finally, $\psi_4$ has complementary functions $\psi_4^\mathrm{CF} \in H$ and particular integral $\psi_4^\mathrm{PI} = - \phi_2 \in L$. Therefore, all together, we have a complementary function which is a sum of `Laplace-like' and `Helmholtz-like' terms:
\begin{equation}
\psi^\mathrm{CF} = \psi^L + \psi^H, \quad \text{ where } \quad \nabla^2 \psi^L = 0, \quad \nabla^2 \psi^H - \psi^H = 0,
\end{equation}
or explicitly, only considering terms that generate a velocity field periodic in $\theta$
\begin{equation}
\begin{split}
\psi^{\mathrm{CF}} =\, A_0 \log r + C_0 \theta + \sum_{n=1}^\infty r^{-n} \left(A_n \cos n\theta + B_n \sin n\theta\right ) + r^{n} \left(C_n \cos n\theta + D_n \sin n\theta\right ) \\
 + E_0 K_0(r) + G_0 I_0(r) + \sum_{n=1}^\infty K_n(r) \left(E_n \cos n\theta + F_n \sin n\theta\right) + I_n(r) \left(G_n \cos n\theta + H_n \sin n\theta\right),
\end{split}
\end{equation}
where the constants $\{A_0, C_0, ..., H_n\}$ are all constants with respect to position and $I_n(r), K_n(r)$ are modified Bessel functions of the first and second kind respectively \cite{lebedev_special_1965}. Automatically, we fix the constants $E_0 = G_0 = 0$, by looking for a solution with zero circulation $\Omega = \int_\mathrm{fluid} \omega_z \dd S = 0$. We also require $A_0 = 0$ to give a pressure that is periodic in $\theta$. To see this, substitute the velocity back into the Stokes equation:
\begin{equation*}
\begin{split}
\nabla p = \nabla^2 \bv{v} - \bv{v} &= \nabla^2 \left[\curl (\psi \uVec_z)\right] - \curl (\psi \uVec_z) = \curl \left[(\nabla^2 \psi - \psi)\uVec_z\right] = - \curl \left(\psi^L \uVec_z\right) \\
 & = \uVec_r \left[ - \frac{C_0}{r} + ... \right] + \uVec_\theta \left[ + \frac{A_0}{r} + ... \right] \;,
\end{split}
\end{equation*}
and integrating the tangential component of this equation: $\frac{1}{r} \partial_\theta p = \frac{A_0}{r} + ... \Rightarrow \, p = A_0 \theta + ...$ which is aperiodic in $\theta$, and hence excluded. Physically relevant solutions to the homogeneous Stokes equation are then of the form:
\begin{subequations}\label{eqn:sComplementaryFunctions}
\begin{gather}
\psi^{\mathrm{CF}} =\, C_0 \theta + \sum_{n=1}^\infty r^{-n} \left(A_n \cos n\theta + B_n \sin n\theta\right ) + r^{n} \left(C_n \cos n\theta + D_n \sin n\theta\right ) \nonumber \\
 + \sum_{n=1}^\infty K_n(r) \left(E_n \cos n\theta + F_n \sin n\theta\right) + I_n(r) \left(G_n \cos n\theta + H_n \sin n\theta\right), \\
p^{\mathrm{CF}} = D_0 - C_0 \log r + \sum_{n=1} r^{-n} \left(- A_n \sin n\theta + B_n \cos n\theta \right) + r^n \left(C_n \sin n\theta - D_n \cos n\theta \right) \;.
\end{gather}
\end{subequations}

The remaining constants are found by substituting the stream function and pressure into each component of the dynamic boundary conditions at each boundary, comparing coefficients of trigonometric functions and solving the resulting linear system. In practice, this step is performed using Mathematica, the code for which we have made available at \href{https://github.com/andra516/dynamicsOfWoundClosure}{https://github.com/andra516/dynamicsOfWoundClosure}. At $\order(1)$ we find (reinstating superscripts):
\begin{gather}
\psi^0 = \tilde{c}_0 \theta, \quad p^0 = \tilde{d_0} - \tilde{c}_0 \log r \quad \Rightarrow \quad \bv{v}^0 = \frac{\tilde{c}_0}{r} \uVec_r,
\end{gather}
where
\begin{equation}
\tilde{c}_0 = - \frac{R_1^0 R_2^0 \left(R_2^0 + R_1^0 \right) + P (R_1^0 R_2^0)^2}{2\left[(R_2^0)^2 - (R_1^0)^2\right] + (R_1^0 R_2^0)^2 \log \frac{R_2^0}{R_1^0}}.
\end{equation}

Finally, the time evolution of each boundary is determined by substituting the velocity field into the kinematic boundary condition~\eqref{eqn:sKBC0}:
\begin{equation*}
\dot{R}_1^0 = \frac{\tilde{c}_0}{R_1^0}, \quad \dot{R}_2^0 = \frac{\tilde{c}_0}{R_2^0} \;,
\end{equation*}
and hence:
\begin{subequations}\label{eqn:sOrder0Shape}
\begin{equation}
\dot{R}_1^0 = - \frac{R_2^0 \left(R_2^0 + R_1^0 \right) + P R_1^0 (R_2^0)^2}{2\left[(R_2^0)^2 - (R_1^0)^2\right] + (R_1^0 R_2^0)^2 \log \frac{R_2^0}{R_1^0}} \;,
\end{equation}
and a similar equation for $\dot{R}_2^0$:
\begin{equation}
\dot{R}_2^0 = - \frac{R_1^0 \left(R_2^0 + R_1^0 \right) + P R_2^0 (R_1^0)^2}{2\left[(R_2^0)^2 - (R_1^0)^2\right] + (R_1^0 R_2^0)^2 \log \frac{R_2^0}{R_1^0}} \;.
\end{equation}
\end{subequations}

In practice, whenever numerically integrating ODEs for the shape modes (such as those in equations~\eqref{eqn:sOrder0Shape}), unless otherwise stated, we take $R_1^0(t=0)=1, R_2^0(t=0)=20$. This ensures the outer boundary remains approximately stationary throughout closure, $\dot{R}^0_2/R^0_2 \overset{\mathrm{incomp.}}{=} \dot{R}^0_1 R^0_1/(R^0_2)^2 \ll 1$.

\subsection{Boundary Conditions at $\order (\pert)$}\label{sec:sBCsAtFirstOrder}
Having solved the passive problem, we now state the boundary conditions on the flow at $\order(\pert)$. Expanding the full boundary conditions and substituting the leading order flow and nematic texture, the $\order(\pert)$ DBCs at the inner boundary read:
\begin{subequations}
\begin{gather}
    -\frac{R_1^1 + \left(R_1^1\right)^{\prime\prime}}{\left(R_1^0\right)^2}=\left[\frac{\tilde{c}_0 R_1^1}{R_1^0}+\frac{4 \tilde{c}_0
   R_1^1}{\left(R_1^0\right)^3}-p^1 +2 \partial_r v_r^1-\beta _1 \right] \bigg |_{R_1^0}, \\
   0=\left[-\frac{4 \tilde{c}_0\left(R_1^1\right)^{\prime}}{\left(R_1^0\right)^3}+\frac{\partial_\theta v_r^1}{R_1^0}-\frac{v_\theta^1}{R_1^0}+\partial_r v_\theta^1\right]\bigg |_{R_1^0},
\end{gather}
and at the outer:
\begin{gather}
    \frac{R_2^1+\left(R_2^1\right)^{\prime\prime}}{\left(R_2^0\right)^2}=\left[\frac{\tilde{c}_0 R_2^1}{R_2^0}+\frac{4 \tilde{c}_0 R_2^1}{\left(R_2^0\right)^3}-p^1 +2 \partial_r v_r^1 +\beta_2 \cos 2 \theta\right] \bigg |_{R_2^0},\\
   0=\left[-\frac{4 \tilde{c}_0 \left(R_2^1\right)^{\prime}}{\left(R_2^0\right)^3}+\frac{\partial_\theta v_r^1}{R_2^0}+\partial_r v_\theta^1-\frac{v_\theta^1}{R_2^0}-\beta_2 \sin 2 \theta\right] \bigg |_{R_2^0},
\end{gather}
\end{subequations}
where $(\cdot)^\prime$ denotes differentiation with respect to $\theta$. The kinematic boundary condition on each boundary at $\order(\pert)$ reads:
\begin{equation}\label{eqn:sKBC1}
v_r^1 |_{R_i^0} = \dot{R}_i^1 + \frac{\tilde{c}_0 R_i^1}{(R_i^0)^2} \;.
\end{equation}

\subsection{Flow at $\order(\pert)$}\label{sec:sFlowAtFirstOrder}
The Stokes equation at $\order(\pert)$ is similar to that of the unperturbed problem, with additional driving from $(\nabla \cdot \bv{Q}^0)$. Therefore, the complementary functions for the stream function and pressure will be given by equations~\eqref{eqn:sComplementaryFunctions}, with additional particular integrals coming from the driving. Taking the curl of the Stokes equation at $\order(\pert)$ and considering the $z$ component, the stream function satisfies:
\begin{equation}
 \left[\curl (\nabla \cdot \bv{Q}^0)\right]_z = \nabla ^4 \psi^1 - \nabla^2 \psi^1 \;,
\end{equation}
where (see eqn. \eqref{eqn:sCurlDivQ0})
\begin{gather*}
\left[\curl \left(\nabla \cdot \bv{Q}^0\right)\right]_z = -\Lambda ^2 \sin 2 \theta \left[f_0 I_2(\Lambda r)+e_0 K_2(\Lambda r)\right] \\
- \sin 4 \theta \left[\Lambda ^2 \left[e_2 K_4(\Lambda r)+g_2 I_4(\Lambda r)\right]-\frac{12 b_2}{r^4}\right].
\end{gather*}
The pressure satisfies:
\begin{equation}
\nabla^2 p^1 = \nabla \cdot (\nabla \cdot \bv{Q}^0).
\end{equation}

Therefore, together with complementary functions given by equations~\eqref{eqn:sComplementaryFunctions}, we also have particular integrals:
\begin{subequations}
\begin{equation}
\psi^1_\mathrm{PI} = \frac{\left[e_0 K_2(\Lambda r) + f_0 I_2(\Lambda r) \right]}{1-\Lambda^2} \sin 2\theta + \left[\frac{b_2}{r^2} + \frac{e_2 K_4(\Lambda r) + g_2 I_4(\Lambda r)}{2(1-\Lambda^2)}\right] \sin 4\theta, 
\end{equation} 
\begin{equation}
\begin{split}
p^1_\mathrm{PI} = d_2 r^2 + \frac{e_2}{2} K_0(\Lambda r) + \frac{g_2}{2} I_0(\Lambda r) + \cos 2\theta \left[e_0 K_2(\Lambda r) + f_0 I_2(\Lambda r)\right] \\ 
 + \cos 4 \theta \left[\frac{12 b_2}{r^4} + \frac{b_2}{r^2} + \frac{e_2}{2} K_4(\Lambda r) + \frac{g}{2} I_4(\Lambda r) \right] \;.
\end{split}
\end{equation}
\end{subequations}

Substituting into the dynamic boundary conditions at each boundary and equating coefficients of trigonometric functions, we have to solve the following linear problems for the constants in the complementary functions (equations~\eqref{eqn:sComplementaryFunctions}):
\begin{subequations}
\begin{equation}
    \begin{pmatrix}
    \mathcal{A}^0_{11}&\mathcal{A}^0_{12} \\
    \mathcal{A}^0_{21}&\mathcal{A}^0_{22} 
    \end{pmatrix}
    \begin{pmatrix}
        \tilde{c}_0 \\ D_0
    \end{pmatrix}
    =
    \begin{pmatrix}
        \mathcal{V}^0_1 \\
        \mathcal{V}^0_2
    \end{pmatrix} \;,
\end{equation}
where
\begin{equation*}
\begin{split}
    &\mathcal{A}_{11}^0 = \frac{2}{\left(R_1^0\right){}^2}-\log \left(R_1^0\right) \;, \\
    &\mathcal{A}_{12}^0 = 1 \;, \\
    &\mathcal{A}_{21}^0 =  \frac{2}{\left(R_2^0\right){}^2}-\log \left(R_2^0\right) \;,\\
    &\mathcal{A}_{22}^0 = 1\;,
\end{split}
\end{equation*}
and 
\begin{equation*}
    \begin{split}
        &\mathcal{V}_1^0 = \frac{\tilde{c}_0 \xi_1^0}{R_1^0}+\frac{4 \tilde{c}_0 \xi_1^0}{\left(R_1^0\right){}^3}-d_2 \left(R_1^0\right){}^2-\frac{1}{2} e_2 K_0\left(\Lambda  R_1^0\right)-\frac{1}{2} g_2 I_0\left(\Lambda  R_1^0\right)+\frac{\xi_1^0}{\left(R_1^0\right){}^2}-\beta _1 \;,\\
        &\mathcal{V}_2^0 = \frac{\xi_2^0 \left(\tilde{c}_0 \left(\left(R_2^0\right){}^2+4\right)-R_2^0\right)}{\left(R_2^0\right){}^3}-d_2 \left(R_2^0\right){}^2-\frac{1}{2} e_2 K_0\left(\Lambda  R_2^0\right)-\frac{1}{2} g_2 I_0\left(\Lambda  R_2^0\right) \;.
    \end{split}
\end{equation*}

\begin{equation}
    \begin{pmatrix}
    \mathcal{P}^2_{11}&\mathcal{P}^2_{12}& \mathcal{P}^2_{13}&\mathcal{P}^2_{14} \\
    \mathcal{P}^2_{11}&\mathcal{P}^2_{12}& \mathcal{P}^2_{13}&\mathcal{P}^2_{14} \\
    \mathcal{P}^2_{11}&\mathcal{P}^2_{12}& \mathcal{P}^2_{13}&\mathcal{P}^2_{14} \\
    \mathcal{P}^2_{11}&\mathcal{P}^2_{12}& \mathcal{P}^2_{13}&\mathcal{P}^2_{14} \\
    \end{pmatrix}
    \begin{pmatrix}
        A_2 \\ C_2 \\ E_2 \\ G_2
    \end{pmatrix}
    =
    \begin{pmatrix}
        \mathcal{W}^2_1 \\
        \mathcal{W}^2_2 \\
        \mathcal{W}^2_3 \\
        \mathcal{W}^2_4 \\
    \end{pmatrix} \;, 
\end{equation}
where
\begin{equation*}
    \begin{split}
        & \mathcal{P}^2_{11} = -\left(R_1^0\right){}^2-12 \;, \\
        & \mathcal{P}^2_{12} =R_1^0 \left(\left(R_1^0\right){}^5+4 \left(R_1^0\right){}^3\right) \;, \\
        & \mathcal{P}^2_{13} = -4 \left(R_1^0\right){}^2 \left(3 K_2\left(R_1^0\right)+R_1^0 K_1\left(R_1^0\right)\right) \;,\\
        & \mathcal{P}^2_{14} = 4 \left(R_1^0\right){}^2 \left(R_1^0 I_1\left(R_1^0\right)-3 I_2\left(R_1^0\right)\right) \;,\\
        & \mathcal{P}^2_{21} = \frac{12}{\left(R_1^0\right){}^4} \;, \\
        & \mathcal{P}^2_{22} =  4\;,\\
        & \mathcal{P}^2_{23} =  \frac{2 K_1\left(R_1^0\right)}{R_1^0}+K_2\left(R_1^0\right)+\frac{12 K_2\left(R_1^0\right)}{\left(R_1^0\right){}^2} \;,\\
        & \mathcal{P}^2_{24} =  -\frac{2 I_1\left(R_1^0\right)}{R_1^0}+I_2\left(R_1^0\right)+\frac{12 I_2\left(R_1^0\right)}{\left(R_1^0\right){}^2}\;, \\
        & \mathcal{P}^2_{31} = -\left(R_2^0\right){}^2-12 \;, \\
        & \mathcal{P}^2_{32} = R_2^0 \left(\left(R_2^0\right){}^5+4 \left(R_2^0\right){}^3\right) \;,\\
        & \mathcal{P}^2_{33} = -4 \left(R_2^0\right){}^2 \left(3 K_2\left(R_2^0\right)+R_2^0 K_1\left(R_2^0\right)\right) \;,\\
        & \mathcal{P}^2_{34} =  4 \left(R_2^0\right){}^2 \left(R_2^0 I_1\left(R_2^0\right)-3 I_2\left(R_2^0\right)\right)\;, \\
        & \mathcal{P}^2_{41} =  \frac{12}{\left(R_2^0\right){}^4} \;,\\
        & \mathcal{P}^2_{42} =  4\;, \\
        & \mathcal{P}^2_{43} = \frac{2 K_1\left(R_2^0\right)}{R_2^0}+K_2\left(R_2^0\right)+\frac{12 K_2\left(R_2^0\right)}{\left(R_2^0\right){}^2}\;, \\
        & \mathcal{P}^2_{44} =  -\frac{2 I_1\left(R_2^0\right)}{R_2^0}+I_2\left(R_2^0\right)+\frac{12 I_2\left(R_2^0\right)}{\left(R_2^0\right){}^2}\;,
    \end{split}
\end{equation*}
and
\begin{equation*}
    \begin{split}
        & \mathcal{W}^2_1 = R_1^0 \eta_1^2 \left(\tilde{c}_0 \left(\left(R_1^0\right){}^2+4\right)-3 R_1^0\right) \;,\\
        & \mathcal{W}^2_2 =  -\frac{8 \tilde{c}_0 \eta_1^2}{\left(R_1^0\right){}^3}\;, \\ 
        & \mathcal{W}^2_3 = R_2^0 \eta_2^2 \left(\tilde{c}_0 \left(\left(R_2^0\right){}^2+4\right)+3 R_2^0\right) \;,\\
        & \mathcal{W}^2_4 = -\frac{8 \tilde{c}_0 \eta_2^2}{\left(R_2^0\right){}^3} \;.
    \end{split}
\end{equation*}
\begin{equation}
    \begin{pmatrix}
    \mathcal{A}^2_{11}&\mathcal{A}^2_{12}& \mathcal{A}^2_{13}&\mathcal{A}^2_{14} \\
    \mathcal{A}^2_{11}&\mathcal{A}^2_{12}& \mathcal{A}^2_{13}&\mathcal{A}^2_{14} \\
    \mathcal{A}^2_{11}&\mathcal{A}^2_{12}& \mathcal{A}^2_{13}&\mathcal{A}^2_{14} \\
    \mathcal{A}^2_{11}&\mathcal{A}^2_{12}& \mathcal{A}^2_{13}&\mathcal{A}^2_{14} \\
    \end{pmatrix}
    \begin{pmatrix}
        B_2 \\ D_2 \\ F_2 \\ H_2
    \end{pmatrix}
    =
    \begin{pmatrix}
        \mathcal{V}^2_1 \\
        \mathcal{V}^2_2 \\
        \mathcal{V}^2_3 \\
        \mathcal{V}^2_4 \\
    \end{pmatrix} \;, 
\end{equation}
where
\begin{equation*}
    \begin{split}
        &\mathcal{A}^2_{11} =  -\frac{\left(\Lambda ^2-1\right) \left(\left(R_1^0\right){}^2+12\right)}{R_1^0}\;,\\
        &\mathcal{A}^2_{12} = -\left(\left(\Lambda ^2-1\right) \left(-\left(R_1^0\right){}^5-4 \left(R_1^0\right){}^3\right)\right) \;,\\
        &\mathcal{A}^2_{13} =-4 \left(\Lambda ^2-1\right) R_1^0 \left(3 K_2\left(R_1^0\right)+R_1^0 K_1\left(R_1^0\right)\right)\;, \\
        &\mathcal{A}^2_{14} = 4 \left(\Lambda ^2-1\right) R_1^0 \left(R_1^0 I_1\left(R_1^0\right)-3 I_2\left(R_1^0\right)\right) \;,\\
        &\mathcal{A}^2_{21} = -\frac{12 \left(\Lambda ^2-1\right)}{R_1^0} \;,\\
        &\mathcal{A}^2_{22} = -4 \left(\Lambda ^2-1\right) \left(R_1^0\right){}^3 \;,\\
        &\mathcal{A}^2_{23} = -\left(\left(\Lambda ^2-1\right) R_1^0 \left(\left(R_1^0\right){}^2 K_2\left(R_1^0\right)+2 R_1^0 K_1\left(R_1^0\right)+12 K_2\left(R_1^0\right)\right)\right) \;,\\
        &\mathcal{A}^2_{24} = -\left(\left(\Lambda ^2-1\right) R_1^0 \left(\left(R_1^0\right){}^2 I_2\left(R_1^0\right)-2 R_1^0 I_1\left(R_1^0\right)+12 I_2\left(R_1^0\right)\right)\right) \;,\\ 
        &\mathcal{A}^2_{31} = \frac{\left(\Lambda ^2-1\right) \left(\left(R_2^0\right){}^2+12\right)}{R_2^0} \;,\\
        &\mathcal{A}^2_{32} = -\left(\left(\Lambda ^2-1\right) \left(R_2^0\right){}^5\right)-4 \left(\Lambda ^2-1\right) \left(R_2^0\right){}^3 \;,\\
        &\mathcal{A}^2_{33} = 4 \left(\Lambda ^2-1\right) R_2^0 \left(3 K_2\left(R_2^0\right)+R_2^0 K_1\left(R_2^0\right)\right) \;,\\
        &\mathcal{A}^2_{34} = -4 \left(\Lambda ^2-1\right) R_2^0 \left(R_2^0 I_1\left(R_2^0\right)-3 I_2\left(R_2^0\right)\right) \;,\\ 
        &\mathcal{A}^2_{41} = -\frac{12 \left(\Lambda ^2-1\right)}{R_2^0} \;,\\
        &\mathcal{A}^2_{42} =  -4 \left(\Lambda ^2-1\right) \left(R_2^0\right){}^3\;,\\
        &\mathcal{A}^2_{43} = -\left(\left(\Lambda ^2-1\right) R_2^0 \left(\left(R_2^0\right){}^2 K_2\left(R_2^0\right)+2 R_2^0 K_1\left(R_2^0\right)+12 K_2\left(R_2^0\right)\right)\right)\;, \\
        &\mathcal{A}^2_{44} = -\left(\left(\Lambda ^2-1\right) R_2^0 \left(\left(R_2^0\right){}^2 I_2\left(R_2^0\right)-2 R_2^0 I_1\left(R_2^0\right)+12 I_2\left(R_2^0\right)\right)\right)\;,
    \end{split}
\end{equation*}
and
\begin{equation*}
    \begin{split}
        &\mathcal{V}^2_1 = -\tilde{c}_0 \Lambda ^2 \left(R_1^0\right){}^2 \xi_1^2+\tilde{c}_0 \left(R_1^0\right){}^2 \xi_1^2-4 \tilde{c}_0 \Lambda ^2 \xi_1^2+4 \tilde{c}_0 \xi_1^2+e_0 \Lambda ^2 \left(R_1^0\right){}^3 K_2\left(\Lambda  R_1^0\right)-e_0 \left(R_1^0\right){}^3 K_2\left(\Lambda  R_1^0\right) \\ & \quad -4 e_0 \Lambda  \left(R_1^0\right){}^2 K_1\left(\Lambda  R_1^0\right)-12 e_0 R_1^0 K_2\left(\Lambda  R_1^0\right)+f_0 \Lambda ^2 \left(R_1^0\right){}^3 I_2\left(\Lambda  R_1^0\right)-f_0 \left(R_1^0\right){}^3 I_2\left(\Lambda  R_1^0\right) \\ & \quad +4 f_0 \Lambda  \left(R_1^0\right){}^2 I_1\left(\Lambda  R_1^0\right)-12 f_0 R_1^0 I_2\left(\Lambda  R_1^0\right)+3 \Lambda ^2 R_1^0 \xi_1^2-3 R_1^0 \xi_1^2 \;,\\
        &\mathcal{V}^2_2 =  -8 \tilde{c}_0 \Lambda ^2 \xi_1^2+8 \tilde{c}_0 \xi_1^2+e_0 \left(-\Lambda ^2\right) \left(R_1^0\right){}^3 K_2\left(\Lambda  R_1^0\right)-2 e_0 \Lambda  \left(R_1^0\right){}^2 K_1\left(\Lambda  R_1^0\right)-12 e_0 R_1^0 K_2\left(\Lambda  R_1^0\right) \\
        & \quad -f_0 \Lambda ^2 \left(R_1^0\right){}^3 I_2\left(\Lambda  R_1^0\right)+2 f_0 \Lambda  \left(R_1^0\right){}^2 I_1\left(\Lambda  R_1^0\right)-12 f_0 R_1^0 I_2\left(\Lambda  R_1^0\right) \;,\\ 
        &\mathcal{V}^2_3 = \tilde{c}_0 \Lambda ^2 \left(R_2^0\right){}^2 \xi_2^2-\tilde{c}_0 \left(R_2^0\right){}^2 \xi_2^2+4 \tilde{c}_0 \Lambda ^2 \xi_2^2-4 \tilde{c}_0 \xi_2^2+e_0 \left(-\Lambda ^2\right) \left(R_2^0\right){}^3 K_2\left(\Lambda  R_2^0\right)+e_0 \left(R_2^0\right){}^3 K_2\left(\Lambda  R_2^0\right) \\ 
        & \quad +4 e_0 \Lambda  \left(R_2^0\right){}^2 K_1\left(\Lambda  R_2^0\right)+12 e_0 R_2^0 K_2\left(\Lambda  R_2^0\right)-f_0 \Lambda ^2 \left(R_2^0\right){}^3 I_2\left(\Lambda  R_2^0\right)+f_0 \left(R_2^0\right){}^3 I_2\left(\Lambda  R_2^0\right) \\
        & \quad -4 f_0 \Lambda  \left(R_2^0\right){}^2 I_1\left(\Lambda  R_2^0\right)+12 f_0 R_2^0 I_2\left(\Lambda  R_2^0\right)+\beta _2 \Lambda ^2 \left(R_2^0\right){}^3-\beta _2 \left(R_2^0\right){}^3+3 \Lambda ^2 R_2^0 \xi_2^2-3 R_2^0 \xi_2^2 \;,\\ 
        &\mathcal{V}^2_4 = -8 \tilde{c}_0 \Lambda ^2 \xi_2^2+8 \tilde{c}_0 \xi_2^2+e_0 \left(-\Lambda ^2\right) \left(R_2^0\right){}^3 K_2\left(\Lambda  R_2^0\right)-2 e_0 \Lambda  \left(R_2^0\right){}^2 K_1\left(\Lambda  R_2^0\right)-12 e_0 R_2^0 K_2\left(\Lambda  R_2^0\right) \\
        & \quad -f_0 \Lambda ^2 \left(R_2^0\right){}^3 I_2\left(\Lambda  R_2^0\right)+2 f_0 \Lambda  \left(R_2^0\right){}^2 I_1\left(\Lambda  R_2^0\right)-12 f_0 R_2^0 I_2\left(\Lambda  R_2^0\right)+\beta _2 \Lambda ^2 \left(R_2^0\right){}^3-\beta _2 \left(R_2^0\right){}^3 \;.
    \end{split}    
\end{equation*}
\begin{equation}
    \begin{pmatrix}
    \mathcal{P}^4_{11}&\mathcal{P}^4_{12}& \mathcal{P}^4_{13}&\mathcal{P}^4_{14} \\
    \mathcal{P}^4_{11}&\mathcal{P}^4_{12}& \mathcal{P}^4_{13}&\mathcal{P}^4_{14} \\
    \mathcal{P}^4_{11}&\mathcal{P}^4_{12}& \mathcal{P}^4_{13}&\mathcal{P}^4_{14} \\
    \mathcal{P}^4_{11}&\mathcal{P}^4_{12}& \mathcal{P}^4_{13}&\mathcal{P}^4_{14} \\
    \end{pmatrix}
    \begin{pmatrix}
        A_4 \\ C_4 \\ E_4 \\ G_4
    \end{pmatrix}
    =
    \begin{pmatrix}
        \mathcal{W}^4_1 \\
        \mathcal{W}^4_2 \\
        \mathcal{W}^4_3 \\
        \mathcal{W}^4_4 \\
    \end{pmatrix} \;, 
\end{equation}
where
\begin{equation*}
    \begin{split}
        &\mathcal{P}^4_{11} = -\left(R_1^0\right){}^2-40 \;,\\
        &\mathcal{P}^4_{12} = \left(R_1^0\right){}^3 \left(\left(R_1^0\right){}^7+24 \left(R_1^0\right){}^5\right)\;,\\
        &\mathcal{P}^4_{13} = -8 \left(R_1^0\right){}^4 \left(5 K_4\left(R_1^0\right)+R_1^0 K_3\left(R_1^0\right)\right) \;,\\
        &\mathcal{P}^4_{14} = 8 \left(R_1^0\right){}^4 \left(R_1^0 I_3\left(R_1^0\right)-5 I_4\left(R_1^0\right)\right) \;,\\ 
        &\mathcal{P}^4_{21} = 40 \;,\\
        &\mathcal{P}^4_{22} = 24 \left(R_1^0\right){}^8 \;,\\
        &\mathcal{P}^4_{23} = \left(R_1^0\right){}^4 \left(\left(R_1^0\right){}^2 K_4\left(R_1^0\right)+2 R_1^0 K_3\left(R_1^0\right)+40 K_4\left(R_1^0\right)\right) \;,\\
        &\mathcal{P}^4_{24} = \left(R_1^0\right){}^4 \left(\left(R_1^0\right){}^2 I_4\left(R_1^0\right)-2 R_1^0 I_3\left(R_1^0\right)+40 I_4\left(R_1^0\right)\right) \;,\\ 
        &\mathcal{P}^4_{31} = -\left(R_2^0\right){}^2-40\;,\\
        &\mathcal{P}^4_{32} = \left(R_2^0\right){}^3 \left(\left(R_2^0\right){}^7+24 \left(R_2^0\right){}^5\right) \;,\\
        &\mathcal{P}^4_{33} = -8 \left(R_2^0\right){}^4 \left(5 K_4\left(R_2^0\right)+R_2^0 K_3\left(R_2^0\right)\right) \;,\\
        &\mathcal{P}^4_{34} = 8 \left(R_2^0\right){}^4 \left(R_2^0 I_3\left(R_2^0\right)-5 I_4\left(R_2^0\right)\right) \;,\\ 
        &\mathcal{P}^4_{41} = 40\;, \\
        &\mathcal{P}^4_{42} =24 \left(R_2^0\right){}^8\;, \\
        &\mathcal{P}^4_{43} = \left(R_2^0\right){}^4 \left(\left(R_2^0\right){}^2 K_4\left(R_2^0\right)+2 R_2^0 K_3\left(R_2^0\right)+40 K_4\left(R_2^0\right)\right) \;,\\
        &\mathcal{P}^4_{44} = \left(R_2^0\right){}^4 \left(\left(R_2^0\right){}^2 I_4\left(R_2^0\right)-2 R_2^0 I_3\left(R_2^0\right)+40 I_4\left(R_2^0\right)\right)\;,
    \end{split}
\end{equation*}
and
\begin{equation}
    \begin{split}
        &\mathcal{W}^4_{1}=\left(R_1^0\right){}^3 \eta_1^4 \left(\tilde{c}_0 \left(\left(R_1^0\right){}^2+4\right)-15 R_1^0\right) \;,\\
        &\mathcal{W}^4_{2}= -16 \tilde{c}_0 \left(R_1^0\right){}^3 \eta_1^4 \;,\\ 
        &\mathcal{W}^4_{3}=\left(R_2^0\right){}^3 \eta_2^4 \left(\tilde{c}_0 \left(\left(R_2^0\right){}^2+4\right)+15 R_2^0\right) \;,\\ 
        &\mathcal{W}^4_{4}=-16 \tilde{c}_0 \left(R_2^0\right){}^3 \eta_2^4\;.
    \end{split}
\end{equation}
\begin{equation}
    \begin{pmatrix}
    \mathcal{A}^4_{11}&\mathcal{A}^4_{12}& \mathcal{A}^4_{13}&\mathcal{A}^4_{14} \\
    \mathcal{A}^4_{11}&\mathcal{A}^4_{12}& \mathcal{A}^4_{13}&\mathcal{A}^4_{14} \\
    \mathcal{A}^4_{11}&\mathcal{A}^4_{12}& \mathcal{A}^4_{13}&\mathcal{A}^4_{14} \\
    \mathcal{A}^4_{11}&\mathcal{A}^4_{12}& \mathcal{A}^4_{13}&\mathcal{A}^4_{14} \\
    \end{pmatrix}
    \begin{pmatrix}
        B_4 \\ D_4 \\ F_4 \\ H_4
    \end{pmatrix}
    =
    \begin{pmatrix}
        \mathcal{V}^4_1 \\
        \mathcal{V}^4_2 \\
        \mathcal{V}^4_3 \\
        \mathcal{V}^4_4 \\
    \end{pmatrix} \;, 
\end{equation}
\end{subequations}
where 
\begin{equation*}
    \begin{split}
        &\mathcal{A}^4_{11} = \frac{\left(R_1^0\right){}^2+40}{\left(R_1^0\right){}^6} \;,\\
        &\mathcal{A}^4_{12} = -\left(R_1^0\right){}^4-24 \left(R_1^0\right){}^2 \;,\\
        &\mathcal{A}^4_{13} =  \frac{8 \left(5 K_4\left(R_1^0\right)+R_1^0 K_3\left(R_1^0\right)\right)}{\left(R_1^0\right){}^2} \;,\\
        &\mathcal{A}^4_{14} =-\frac{8 \left(R_1^0 I_3\left(R_1^0\right)-5 I_4\left(R_1^0\right)\right)}{\left(R_1^0\right){}^2} \;,\\ 
        &\mathcal{A}^4_{21} = 80 \;,\\
        &\mathcal{A}^4_{22} = 48 \left(R_1^0\right){}^8 \;,\\
        &\mathcal{A}^4_{23} =  2 \left(R_1^0\right){}^4 \left(\left(R_1^0\right){}^2 K_4\left(R_1^0\right)+2 R_1^0 K_3\left(R_1^0\right)+40 K_4\left(R_1^0\right)\right) \;, \\
        &\mathcal{A}^4_{24} = 2 (R_1^0)^6 I_4(R_1^0) - 4 (R_1^0)^5 I_3(R_1^0) + 80 (R_1^0)^4 I_4(R_1^0)\;, \\
        &\mathcal{A}^4_{31} = \frac{\left(R_2^0\right){}^2+40}{\left(R_2^0\right){}^6} \;,\\
        &\mathcal{A}^4_{32} =  -\left(R_2^0\right){}^4-24 \left(R_2^0\right){}^2 \;,\\
        &\mathcal{A}^4_{33} =  \frac{8 \left(5 K_4\left(R_2^0\right)+R_2^0 K_3\left(R_2^0\right)\right)}{\left(R_2^0\right){}^2} \;,\\
        &\mathcal{A}^4_{34} =  -\frac{8 \left(R_2^0 I_3\left(R_2^0\right)-5 I_4\left(R_2^0\right)\right)}{\left(R_2^0\right){}^2} \;,\\
        &\mathcal{A}^4_{41} = 80 \;,\\
        &\mathcal{A}^4_{42} = 48 (R_2^0)^8 \;,\\
        &\mathcal{A}^4_{43} = 2 \left(R_2^0\right){}^4 \left(\left(R_2^0\right){}^2 K_4\left(R_2^0\right)+2 R_2^0 K_3\left(R_2^0\right)+40 K_4\left(R_2^0\right)\right) \;,\\
        &\mathcal{A}^4_{44} = 2 (R_2^0)^6 I_4(R_2^0) - 4 (R_2^0)^5 I_3(R_2^0) + 80 (R_2^0)^4 I_4(R_2^0) \;,
    \end{split}
\end{equation*}
and
\begin{equation*}
    \begin{split}
        &\mathcal{V}^4_1 = -\frac{b_2}{(R_1^0)^2} - \frac{36b_2}{(R_1^0)^4} + \frac{\tilde{c}_0 \xi_1^4}{R_1^0} + \frac{4 \tilde{c}_0 \xi_1^4}{(R_1^0)^3} - \frac{e_2 K_4(\Lambda R_1^0)}{2} + \frac{4 e_2 \Lambda K_3(\Lambda R_1^0)}{(\Lambda^2 - 1) R_1^0} + \frac{20 e_2 K_4(\Lambda R_1^0)}{(\Lambda^2 -1) (R_1^0)^2} \\
        & \quad - \frac{g_2 I_4(\Lambda R_1^0)}{2} - \frac{4g_2 \Lambda I_3(\Lambda R_1^0)}{(\Lambda^2-1)R_1^0} + \frac{20 g_2 I_4(\Lambda R_1^0)}{(\Lambda^2-1) (R_1^0)^2} - \frac{15 \xi_1^4}{(R_1^0)^2}  \;,\\
        &\mathcal{V}^4_2 = -48b_2 (R_1^0)^2 + 32 \tilde{c}_0 (R_1^0)^3 \xi_1^4 + \frac{\Lambda^2 (R_1^0)^6 }{\Lambda^2-1}\left(e_2 K_4(\Lambda R_1^0) + g_2 I_4(\Lambda R_1^0) \right) \\
        & \quad +\frac{2 \Lambda (R_1^0)^5}{\Lambda^2 -1} \left(e_2 K_3(\Lambda R_1^0) - g_2 I_3(\Lambda R_1^0)\right) + \frac{40 (R_1^0)^4}{\Lambda^2-1} \left(e_2 K_4(\Lambda R_1^0) + g_2 I_4(\Lambda R_1^0) \right)\;, \\
        &\mathcal{V}^4_3 = -\frac{b_2}{(R_2^0)^2} - \frac{36b_2}{(R_2^0)^4} + \frac{\tilde{c}_0 \xi_2^4}{R_2^0} + \frac{4 \tilde{c}_0 \xi_2^4}{(R_2^0)^3} - \frac{e_2 K_4(\Lambda R_2^0)}{2} + \frac{4 e_2 \Lambda K_3(\Lambda R_2^0)}{(\Lambda^2 - 1) R_2^0} + \frac{20 e_2 K_4(\Lambda R_2^0)}{(\Lambda^2 -1) (R_2^0)^2} \\
        &\quad - \frac{g_2 I_4(\Lambda R_2^0)}{2} - \frac{4g_2 \Lambda I_3(\Lambda R_2^0)}{(\Lambda^2-1)R_2^0} + \frac{20 g_2 I_4(\Lambda R_2^0)}{(\Lambda^2-1) (R_2^0)^2} - \frac{15 \xi_2^4}{(R_2^0)^2}\;, \\
        &\mathcal{V}^4_4 = -48b_2 (R_2^0)^2 + 32 \tilde{c}_0 (R_2^0)^3 \xi_2^4 + \frac{\Lambda^2 (R_2^0)^6}{\Lambda^2-1} \left(e_2 K_4(\Lambda R_2^0) + g_2 I_4(\Lambda R_2^0) \right) \\
        & \quad + \frac{2 \Lambda (R_2^0)^5}{\Lambda^2-1} \left(e_2 K_3(\Lambda R_2^0) - g_2 I_3(\Lambda R_2^0)\right) + \frac{40 (R_2^0)^4}{\Lambda^2-1} \left(e_2 K_4(\Lambda R_2^0) + g_2 I_4(\Lambda R_2^0) \right)\;.
    \end{split}
\end{equation*}

The exact form of constants $\{C_0, D_0, A_2, ...\}$ are complicated and refer the reader to our \href{https://github.com/andra516/dynamicsOfWoundClosure}{\underline{code}} that solves the problem in full.

\begin{figure*}
    \includegraphics[width=0.8\textwidth]{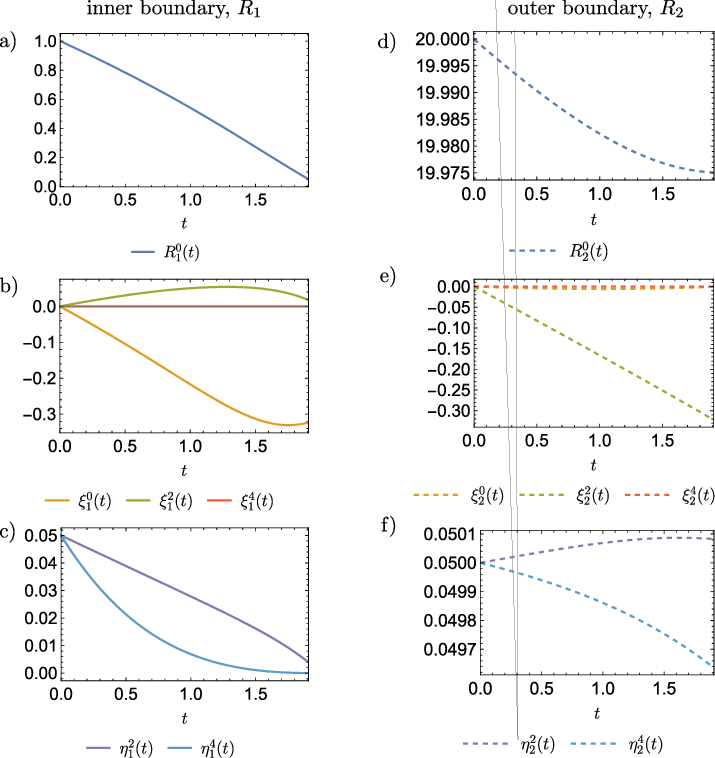}
    \caption{\raggedright Example of shape mode dynamics for the inner boundary, $R_1$ (a-c) and outer boundary $R_2$ (d-f). Plotted using $\Lambda=0.1$ with initial conditions $R_1^0(0)=1, R_2^0(0)=20, \xi_i^k(0)=0 \forall i, k$ and $\eta_i^k(0)=0.05 \forall i,k $.}
    \label{fig:sDetailedShapeDynamics}
\end{figure*}
Finally, we substitute the velocity field into the KBCs at $\order (\pert)$, equation ~\eqref{eqn:sKBC1}. Equating coefficients of trigonometric functions, we obtain ODEs for the shape mode amplitudes, $\dot{\xi}_i^n, \dot{\eta}_i^n$. These ODEs have complex dependencies on $R_1^0, R_2^0, \Lambda$, (see \href{https://github.com/andra516/dynamicsOfWoundClosure}{code} for explicit expressions) however ultimately reduce to expressions of the form:
\begin{subequations}\label{eqn:sShapeModeDynamics}
\begin{equation}
\label{eqn:sInnerShapeModeDynamics}
\begin{split}
	\dot{\xi}_1^0 &= -\Delta_1^0 - a_1^0 \xi_1^0, \\
	\dot{\xi}_1^2 &= \Delta_1^2 - a_1^2 \xi_1^2 + b_1^2 \xi_2^2, \\
	\dot{\xi}_1^4 &= \Delta_1^4  - a_1^4 \xi_1^4 - b_1^4 \xi_2^4 \\
	\dot{\eta}_1^2 &= -c_1^2 \eta_1^2 + d_1^2 \eta_2^2, \\
	\dot{\eta}_1^4 &= -c_1^4 \eta_1^4 - d_1^4 \eta_2^4, \\
\end{split}
\end{equation}
for the inner boundary and
\begin{equation}
\label{eqn:sOuterShapeModeDynamics}
\begin{split}
	\dot{\xi}_2^0 &= -\Delta_2^0 - a_2^0 \xi_1^0 + b_2^0 \xi_2^0, \\
	\dot{\xi}_2^2 &= - \Delta_2^2 + a_2^2 \xi_1^2 - b_2^2 \xi_2^2, \\
	\dot{\xi}_2^4 &= \Delta_2^4  + a_2^4 \xi_1^4 + b_2^4 \xi_2^4 \\
	\dot{\eta}_2^2 &= c_2^2 \eta_1^2 - d_2^2 \eta_2^2, \\
	\dot{\eta}_2^4 &= c_2^4 \eta_1^4 + d_2^4 \eta_2^4, \\
\end{split}
\end{equation}
\end{subequations}
for the outer. Terms $\Delta_{-}^{-}, a_{-}^{-}, b_{-}^{-}, c_{-}^{-}, d_{-}^{-} >0 $ are all complicated functions of $R_1^0, R_2^0, \Lambda$. All modes have relaxational terms arising due to surface tension, however the $\xi_i^{0,2,4}$ modes are driven, as a consequence of activity in the bulk. These ODEs are integrated numerically to determine the shapes of the free boundaries as a function of time. Once the free boundary shape is known, the velocity, pressure and nematic texture fields are also determined, and the problem is solved. Figure~\ref{fig:sDetailedShapeDynamics} shows the typical time evolution of the shape mode amplitudes as a function of time during closure. We use $\beta_1=\beta_2=P=1$ and stop integrating once $R_1^0$ drops below 5\% of its initial value. 

\vspace{0.25cm}

Referring back to equations~\eqref{eqn:sModelWoundQAngle}, we remind the reader that the area, anisotropy and angle of major axis of the shape enclosed by the inner `wound' free surface  are expressed in terms of these shape modes as:
\begin{subequations}
\begin{gather}
    A_\mathrm{w}/\pi = (R_1^0)^2 + 2 \pert \xi_1^0 R_1^0 + \order( \pert^2) \\
    ||q_\mathrm{w}|| = \frac{\sqrt{(\xi_1^2)^2 + (\eta_1^2)^2}}{2\pi R_1^0} |\pert| + O(\pert^2) \\
    \phi_\mathrm{wound} = \frac{1}{2} \arctan \left(\frac{\eta_1^2}{\mathrm{sgn}[\pert]\xi_1^2}\right ) + \order(\pert)
\end{gather}
\end{subequations}

Therefore, since we have $\xi_1^0 < 0$ and $\xi_1^2>0$ throughout closure (see Fig.~\ref{fig:sDetailedShapeDynamics}), we conclude that the effect of \textit{contractile} ($\pert>0$) active bulk stresses are to (a) accelerate wound closure ($\pert \xi_1^0 R_1^0 < 0 $ throughout) and (b) drive the wound away from circular ($||q_\mathrm{w}|| >0$) and (c) alignment is along the $x$ rather than the $y$-axis (since $\mathrm{sgn}[\pert]\xi_1^2 >0$ throughout) -- i.e. the wound anisotropy is aligned along the axis of nematic order in the far-field. This is illustrated more clearly in Figure 2c.

\vspace{0.5cm}

\begin{figure}
    \includegraphics[width=0.5\textwidth]{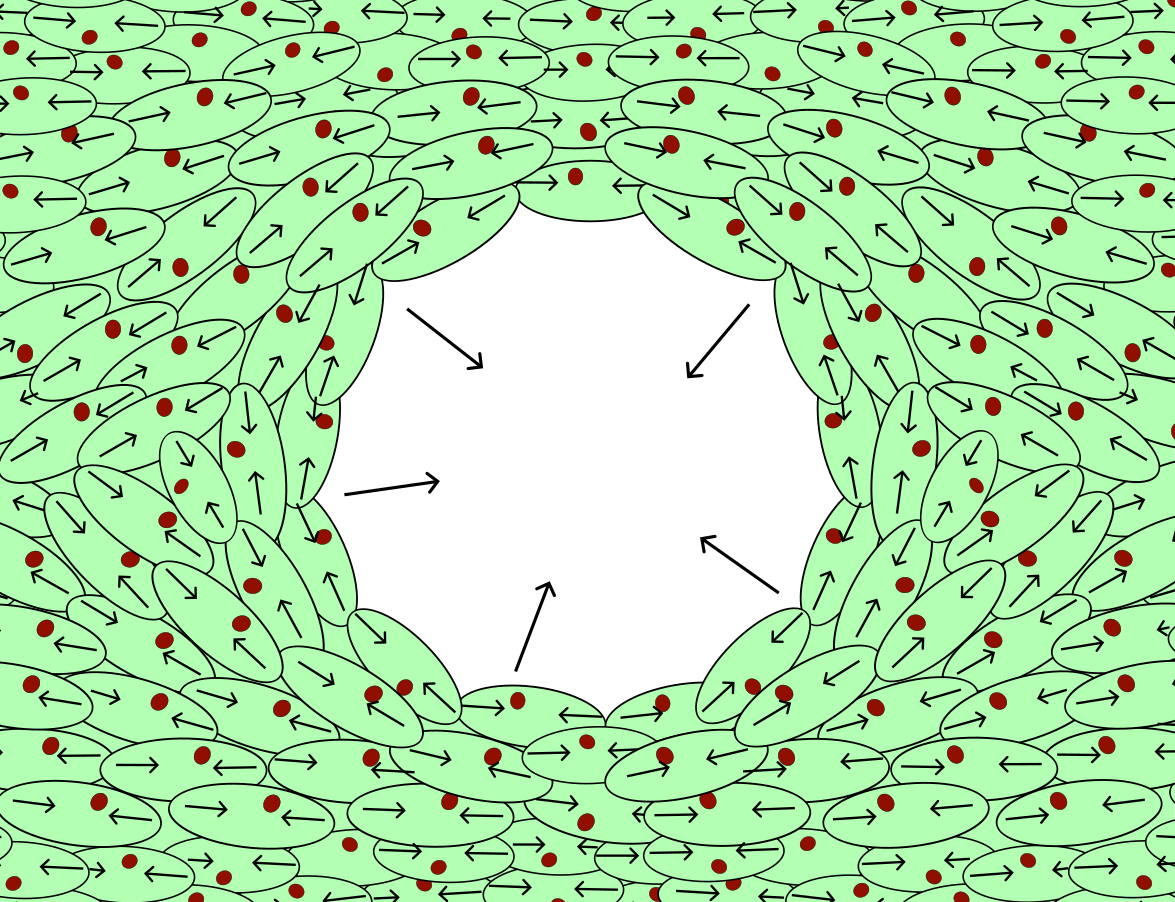}
    \caption{Cartoon illustrating effect of many contractile force dipoles acting along the length of each nematogen/cell, representing the state of active stress in the tissue at the microscopic level.}
    \label{fig:sCartoon}
\end{figure}

Figure~\ref{fig:sCartoon} features a cartoon illustrating why the combination of contractile active stresses and parallel anchoring gives rise to accelerated healing. An active nematic stress of the form $\alpha \bv{Q}$ -- at the continuum level -- may be derived by coarse-graining the effect of many force dipoles acting along the length of each nematogen \cite{simha_hydrodynamic_2002} -- in our case the cells in the epithelium. Aligning each of the cells parallel to the wound boundary and allowing each of them to impose an inwardly pointing (contractile) force dipole along their length results in an inwardly pointing net force at each point, resulting in accelerated closure.

\section{Supplementary Results}
\subsection{Inner Boundary Shape Dynamics are Insensitive to Presence of Outer Boundary}\label{sec:sR2Sensitivity}
\begin{figure*}
\includegraphics[width=0.6\textwidth]{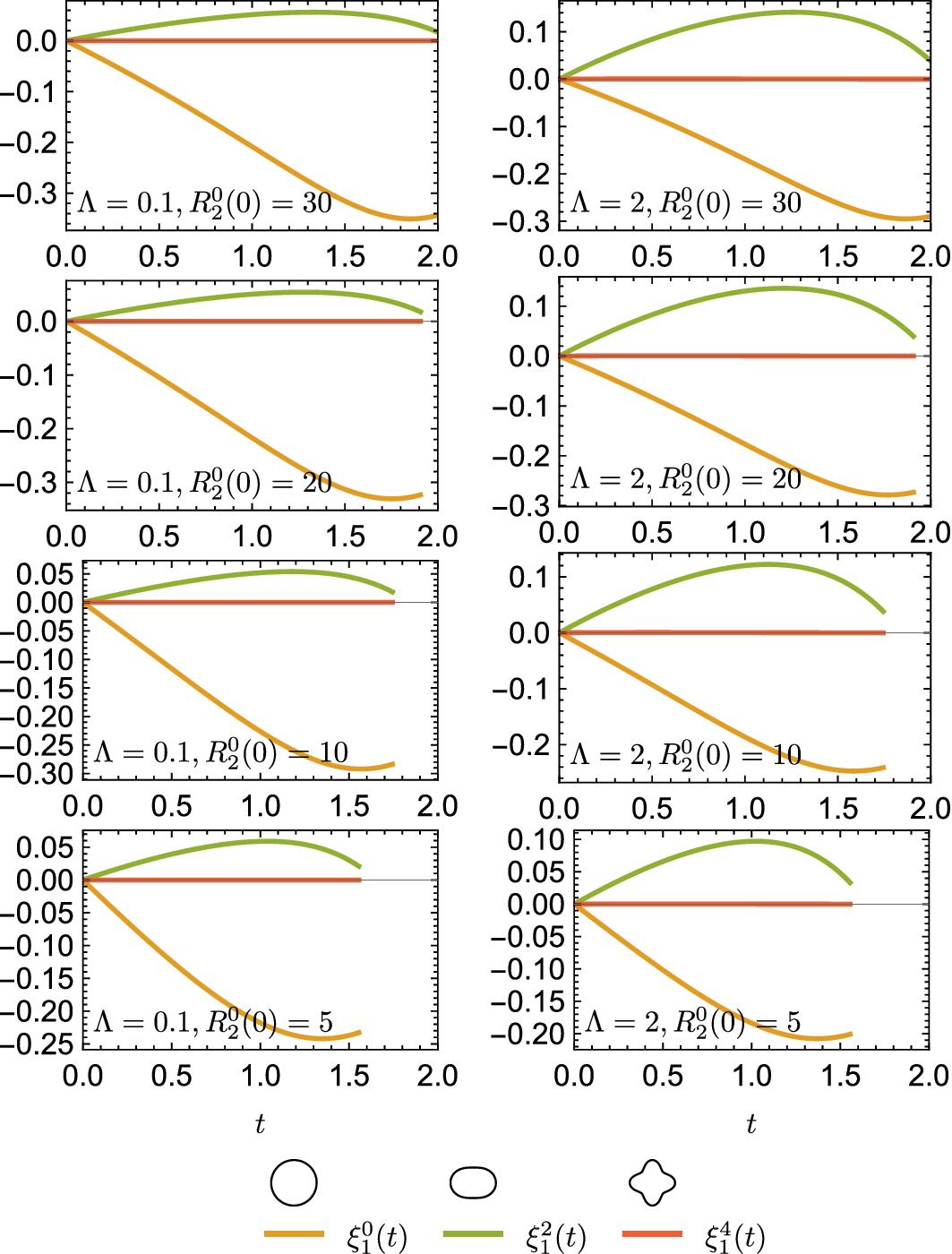}
    \caption{\raggedright Inner boundary shape mode dynamics for $\Lambda=0.1$ (left) and $\Lambda = 2$ (right) for decreasing initial outer boundary radius $R_2^0(0)$. We observe little qualitative difference in the shape dynamics of the inner boundary as the outer boundary radius increases.}
\label{fig:sR2Sensitivity}
\end{figure*}
To investigate the effect of the outer boundary on our conclusions, we integrated ODEs~\eqref{eqn:sShapeModeDynamics} for two values of $\Lambda = 0.1,\, 2$ and with different initial outer boundary radii, from $R_2^0(0) = 30$ down to $R_2^0(0)=5$, fixing the inner boundary initial radius at $R_1^0(0)=1$. We stopped integrating once $R_1^0 \leq 0.01$. Figure~\ref{fig:sR2Sensitivity} shows the time evolution of shape modes for the inner boundary in each case.

For each value of $\Lambda$, the inner hole closed faster as the initial outer boundary radius decreased. Otherwise, there was little qualitative difference in the inner boundary shape dynamics as the outer radius was varied.

\subsection{Effect of Nematic Length Scale on Wound Anisotropy}
\begin{figure*}
\centering
\includegraphics[width=0.9\textwidth]{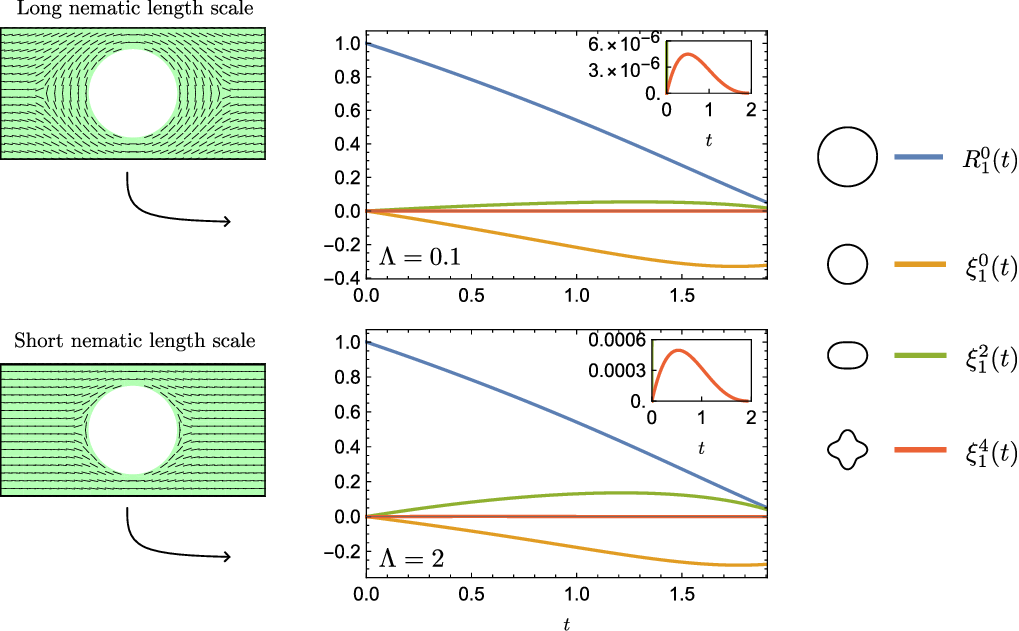}
    \caption{\raggedright Comparison of inner boundary shape evolution $R_1(\theta,t) = R_1^0 + \pert \left(\xi_1^0 + \xi_1^2 \cos 2\theta + \xi_1^4 \cos 4\theta \right)$ for two values of $\Lambda = L / \ell_Q$. We find greater anisotropy in the shape of the inner boundary (greater shape mode amplitudes) as the nematic length scale is decreased.}
\label{fig:sComparingModeDynamics}
\end{figure*}

Comparing the shape dynamics between values of $\Lambda$ (Figure~\ref{fig:sComparingModeDynamics}), we find that the shape mode amplitudes $\xi_1^{2,4}(t)$ generally attain greater maximum values for larger values of $\Lambda$ (i.e. shorter nematic length scales $\sim 1/\Lambda$). That is, the wound boundary becomes more \textit{anisotropic} as the nematic length scale decreases and the active bulk force $(\nabla \cdot \bv{Q}^0)$ is increasingly localised on the boundary. Examining the upper panels of Figure~\ref{fig:sR2Sensitivity}, we see that this result isn't special for this particular choice of the outer boundary initial radius. 

Our model suggests there are two ways of driving anisotropy in the wound shape: (1) by increasing the value of $\pert = \alpha Q_0/\Pi$ (larger $\alpha$ or nematic order $Q_0$ or weaker surface tension, $\Pi=\gamma/L$) or (b) increasing $\Lambda = L/\ell_Q$ (i.e. decreasing the nematic length scale, $\ell_Q$ relative to the `flow' length scale $L^2 = \eta/\Gamma$). Measuring the active stress -- i.e. quantifying $\alpha Q_0$ -- \textit{in vivo}, is generally difficult. However, experimentally measuring the the correlations in cellular alignments (which acts as a proxy for the nematic length scale) would, in principle, allow us to validate the prediction that longer nematic length scales are correlated with weaker non-circularity. It may also allow us estimate (or at least bound) the magnitude of the active nematic stresses in the tissue. This will be the focus of future work.

\subsection{Alternative nematic anchoring on inner boundary}
To investigate the effect of our chosen boundary conditions on closure, we repeated the above calculation with normal anchored boundary conditions on the nematic texture at the inner wound boundary. The calculation follows the same steps as outlined above, except with normal anchoring on the nematic texture at $R_1^0$:
\begin{subequations}
\begin{gather}
    Q_{xx}|_{R_1^0} = \beta_1 \cos 2\theta , \\
    Q_{xy}|_{R_1^0} = \beta_1 \sin 2\theta .    
\end{gather}
\end{subequations}

\begin{figure*}
\centering
\includegraphics[width=0.9\textwidth]{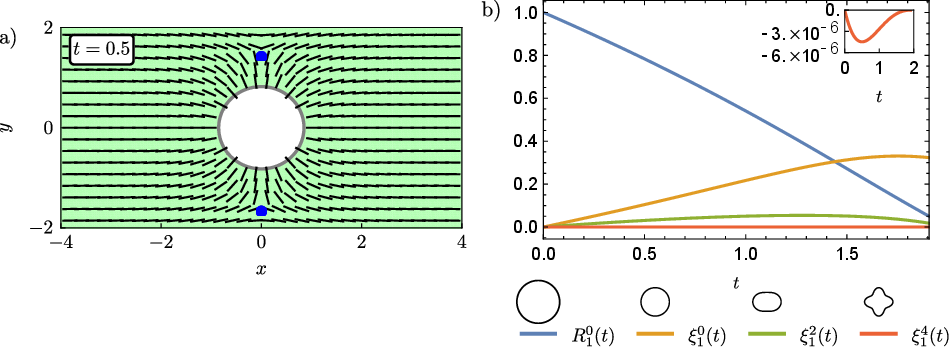}
    \caption{\raggedright (a) Illustration of the nematic texture surrounding the wound boundary with normal anchored boundary conditions on $\bv{Q}^0$. (b) Inner boundary shape mode dynamics. Contrary to the parallel anchored case, contractile ($\alpha>0$) active stresses decelerate closure. Plotted using $\Lambda=0.1$.}
\label{fig:sNormalAnchored}
\end{figure*}

Figure~\ref{fig:sNormalAnchored} illustrates the nematic texture surrounding the wound free boundary for $\Lambda=0.1$ and the time evolution of the driven shape modes.

First, we note the positions of the two $-1/2$ defects now lie along the vertical $y=0$ centreline. Second, the isotropic driven mode $\xi_1^0(t)$ is positive throughout closure. As such, contrary to the parallel anchored case, normal anchoring at the inner boundary requires \textit{extensile} ($\alpha<0$) active stresses to accelerate closure.

\end{document}